\documentclass[12pt]{iopart}
\pdfoutput=1

\usepackage{epsfig,color}
\usepackage{subfig}
\usepackage{verbatim}
\usepackage{graphicx}
\usepackage{latexsym}
\usepackage{amsfonts}
\usepackage{amssymb}
\usepackage{mathrsfs}
\usepackage{bm}

\newcommand{\FR}{\mathbb{R}} 

\newcommand{\Kt}{\mathcal{K}} 

\renewcommand{\rme}{{\mathrm{e}}} 
\renewcommand{\rmi}{{\mathrm{i}}} 
\renewcommand{\rmd}{{\mathrm{d}}} 

\DeclareMathAlphabet{\bi}{OML}{cmm}{b}{it}
\DeclareMathAlphabet{\bcal}{OMS}{cmsy}{b}{n}
\DeclareMathAlphabet{\brmn}{OT1}{cmr}{bx}{n}

\DeclareMathSymbol{\R}{\mathalpha}{AMSb}{"52}

\newcommand{\bgamma}{\bm{\gamma}}

\newcommand{\eqref}[1]{(\ref{#1})}

\def \x{\mathbf{x}}

\def \z{\mathbf{z}}
\def \v{\mathbf{v}}

\def \P{\mathbf{P}}

\def \0{\mathbf{0}}

\def \S{\mathbf{S}}
\def \R{\mathbf{R}}

\def \b{\mathbf{b}}

\def \l{\mathbf{l}}
\def \L{\mathbf{L}}

\begin{document}
\title[Doppler Synthetic Aperture Radar Interferometry]
{Doppler Synthetic Aperture Radar Interferometry: A Novel SAR Interferometry for Height Mapping using Ultra-Narrowband Waveforms}
\author{Birsen Yaz{\i}c{\i}$^{1,*}$, Il-Young Son$^1$ and H. Cagri Yanik}
\address{$^1$Department of Electrical and Computer Systems Engineering, Rensselaer Polytechnic
  Institute, Troy, NY, USA}
\address{$^*$Corresponding author}
\ead{yazici@ecse.rpi.edu}



\begin{abstract}
This paper introduces a new and novel radar interferometry based on Doppler synthetic aperture radar (Doppler-SAR) paradigm. Conventional SAR interferometry relies on wideband transmitted waveforms to obtain high range resolution. Topography of a surface is directly related to the range difference between two antennas configured at different positions.  Doppler-SAR is a novel imaging modality that uses ultra-narrowband continuous waves (UNCW). It takes advantage of high resolution Doppler information provided by UNCWs to form high resolution SAR images.

We introduced the theory of Doppler-SAR interferometry. We derived interferometric phase model and develop the equations of height mapping. Unlike conventional SAR interferometry, we show that the topography of a scene is related to the difference in Doppler between two antennas configured at different velocities. While the conventional SAR interferometry uses range, Doppler and Doppler due to interferometric phase in height mapping, Doppler-SAR interferometry uses Doppler, Doppler-rate and Doppler-rate due to interferometric phase in height mapping. We demonstrate our theory in numerical simulations.

Doppler-SAR interferometry offers the advantages of long-range, robust, environmentally friendly operations; low-power, low-cost, lightweight systems suitable for low-payload platforms, such as micro-satellites; and passive applications using sources of opportunity transmitting UNCW.


\end{abstract}
\submitto{\IP}
\maketitle

\section{Introduction}

Synthetic Aperture Radar (SAR) interferometry is a powerful tool in mapping surface topography and monitoring dynamic processes. This tool is now an integral part of  wide range of applications in many disciplines including environmental remote sensing, geosciences and climate research, earthquake and volcanic research, mapping of Earth's topography, ocean surface current monitoring, hazard and disaster monitoring, as well as defense and security related research \cite{bamler1998synthetic}.

Basic principles of SAR interferometry were originally developed in radio astronomy~\cite{rogers1969venus,rogers1972topography}. Interferometric processing techniques  and systems were later developed and applied to Earth observation~\cite{graham1974synthetic, zebker1986topographic, goldstein1987interferometric, gabriel1988crossed, gabriel1989mapping}.

SAR interferometry exploits phase differences of two or more SAR images to extract more information about a medium than present in a single SAR image \cite{hanssen2001radar} \cite{rosen2000synthetic}. Conventional SAR interferometry relies on wideband transmitted waveforms to obtain high range resolution \cite{rosen2000synthetic, bamler1998synthetic, cherniakov2008bistatic,fritz2011interferometric,duque2010single}. The phase difference of two wideband SAR images are related to range difference. There are many different interferometric methods depending on the configuration of imaging parameters in space, time, frequency etc \cite{bamler1998synthetic}.  When two images are acquired from different look-directions, the phase difference is related to the topography of a surface.

In this paper, we develop the basic principles of a new and novel interferometric method based on Doppler-SAR paradigm to determine topography of a surface. Unlike conventional SAR, Doppler-SAR uses ultra-narrowband continuous waves (CW) to form high resolution images \cite{wang2012bistatic,wang2013ground,wang2013bistatic,wang2012synthetic,wang2012detection,5960500,wang2011ultranarrow, dsah,dsah2,borden2005synthetic,wang2013theory}. Conventional SAR takes advantage of high range resolution and range-rate due to the movement of SAR antenna for high resolution imaging. Doppler-SAR, on the other hand, takes advantage of high temporal Doppler resolution provided by UNCWs and Doppler-rate for high resolution imaging.

We develop the phase relationship between two Doppler-SAR images and show that the phase difference is related to Doppler difference. We approximate this phase difference as Doppler-rate and derive the equations of height mapping for Doppler-SAR interferometry.

Conventional wideband SAR interferometry for height mapping requires two different look-directions. Doppler-SAR interferometry provides a new degree of freedom in system design by allowing antennas to have the same look-direction, but different velocities to obtain height mapping. Additional advantages of Doppler-SAR interferometry include the following: (\emph{i}) Small, lightweight, inexpensive, easy-to-design and calibrate hardware, high Signal-to-Noise-Ratio(SNR) and long effective range of operation.
All of these make Doppler-SAR interferometry a suitable modality for applications requiring high SNR, long range of operation and low payload platforms such as micro-satellites or small uninhabited aerial vehicles. (\emph{ii}) Effective use of electromagnetic spectrum and environmentally friendly illumination. (\emph{iii}) Passive applications. Doppler-SAR may not require dedicated transmitters, since existing Radio Frequency (RF) signals of opportunity often have the ultra-narrowband properties.

To the best of our knowledge, this is the first interferometric method that is developed in Doppler-SAR paradigm. We present the theory for two monostatic Doppler-SAR. However, the method can be easily be extended to bistatic and multi-static configurations and synthetic aperture imaging applications in acoustics.

	The rest of the paper is organized as follows: In Section \ref{sec:Config}, SAR geometry
        and notation are defined. In Section \ref{sec:WidebandInterferometry}, wideband SAR image
        formation, layover effect and basic principles of wideband SAR interferometry are described
        in a perspective relevant to our subsequent development. In Section \ref{sec:Doppler-SAR},
        Doppler-SAR data model, image formation and layover are summarized. Section
        \ref{sec:Doppler-SAR_interferometry} introduces the basic principles of Doppler-SAR
        interferometry and compares the results to wideband SAR case.
        Section~\ref{sec:simulations} presents numerical simulations and
        Section \ref{sec:conclusion} concludes the paper.


\section{Configurations and Notation}
\label{sec:Config}
\begin{figure}[htpb]
\begin{center}
{\includegraphics[width=4in]
          {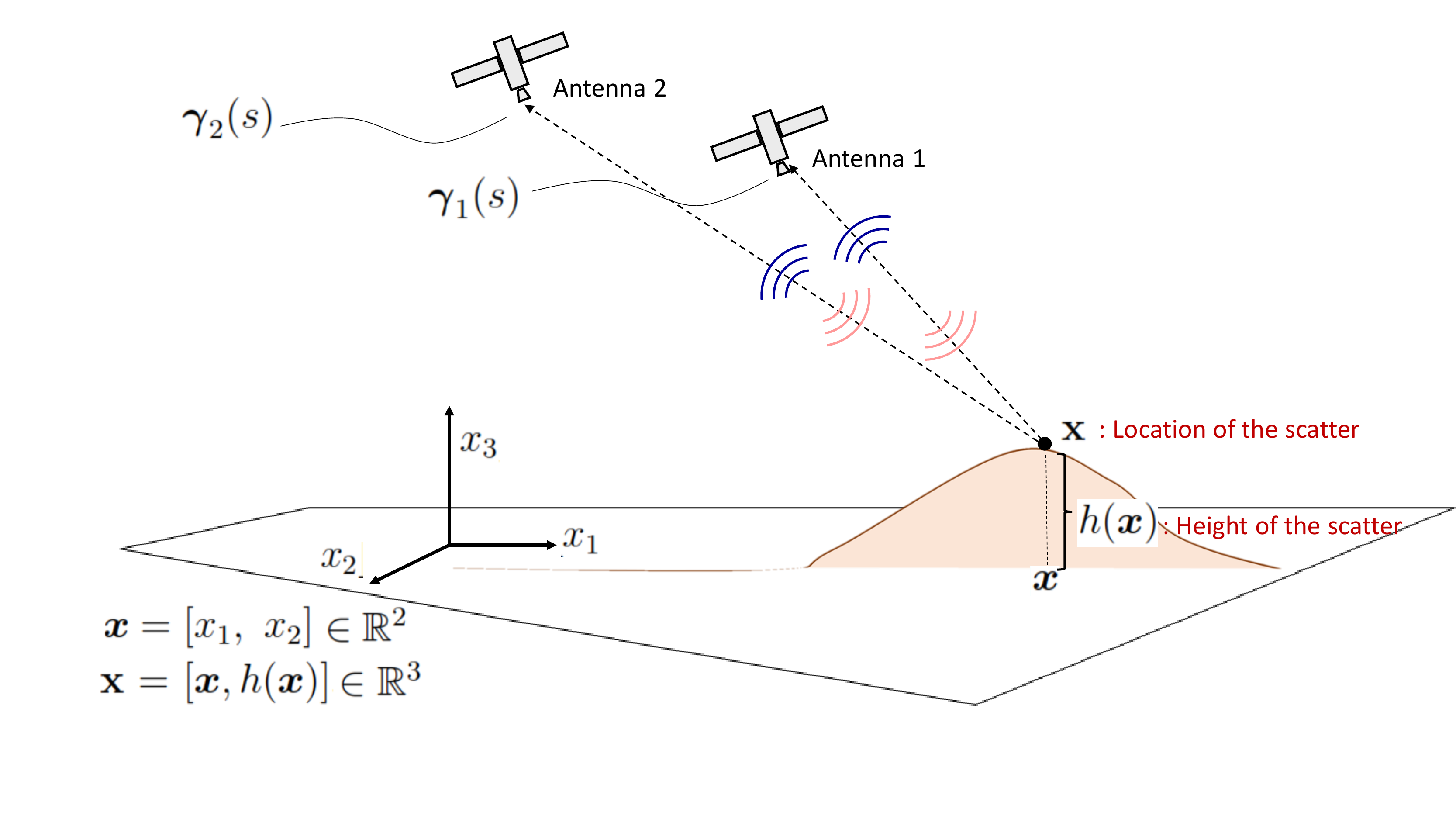}      }
\end{center}
 \caption{Imaging geometry for an interferometric SAR system with two antennas following trajectories $\bgamma_{1}(s)$ and $\bgamma_{2}(s)$. The scatterer is located at $\x \in \FR^3$ where its height is $h(\bi x)$ and $\bi x = [x_1, x_2] \in \FR^2$.}
\label{fig:wb-geo}
\end{figure}

We consider two mono-static SAR systems as shown in Fig. \ref{fig:wb-geo}.

Let $\bgamma_1(s)$ and $\bgamma_2(s)$, $s\in [S_1, \ S_2]\subseteq\mathbb{R}$,  denote the trajectories of the first and second antennas, respectively. 

Unless otherwise stated, bold Roman, bold italic, and Roman
lower-case letters will denote elements in $\FR^3$, $\FR^2$ and $\FR$,
respectively, i.e., $\bi x = [x_1, \ x_2] \in
\FR^2$, $x_3\in \FR$, and $\brmn x=[\bi x,x_3] \in \FR^3$.
The Earth's surface is located at $\brmn x
=[\bi x,h(\bi x)]$, where $h:\FR^2 \rightarrow \FR$, is the
\emph{unknown height} representing ground topography. Let $V:\FR^3 \rightarrow \FR$ denote target reflectivity where we assume that the scattering takes place only on the surface of the Earth. Major notation used throughout the paper is tabulated in Table 1.

\begin{table}
  \caption{Notation}
  \centering 
  {\footnotesize
    \begin{tabular}{p{1.3in}p{3.5in}}
      \hline \hline
      Symbol & Description \\ 
      \hline 
      $\brmn x = [\bi x, h(\bi x)]$, $\brmn x\in\FR^2$ & Location on earth's surface \\
      $h(\bi x)$ & Unknown height of a scatter at $\x$ \\
      $V(\brmn x)$ & Surface reflectivity \\
      $\bgamma_i(s)$ & $i$-th antenna trajectory \\
      $s$ & Slow-time \\
      $t$ & Fast-time \\
      $R_i(\brmn x,s)$ & Range of $i$-th antenna \\
      $\omega_0$ & Center frequency of the transmitted waveforms \\
      $s_0^i$ & Zero-Doppler time for the $i$-th antenna \\
      $\brmn L_i(\brmn x, s)$ & Look-direction of the $i$-th antenna \\
      $d_i^{WB}(t,s)$ & Wideband SAR demodulated received signal at $i$-th antenna \\
      $|\z - \bgamma_i(s)| = \mathrm{C.}$ & Iso-range surface \\
      $\widehat{(\z-\bgamma_i(s))}\cdot\dot{\bgamma}_i(s) = \mathrm{C}$ & Iso-Doppler surface \\
      $\Kt_i^{WB}$ & Filtered backprojection (FBP) operator for wideband SAR \\
      $I_i^{WB}$ & Wideband SAR image \\
      $\Phi^{WB}_{s_0}(\x)$ & Wideband interferometric phase \\
      $\brmn b$ & Baseline vector in wideband SAR interferometry\\
      $\brmn L_1(\brmn z, s_0^1)\cdot\brmn b = \textmd{C}$ & Interferometric phase cone \\
      $\brmn l$ & Vector from a known scatterer position to the unknown location of a scatterer \\
      $\l_1^\bot$ & Component of $\brmn l$ perpendicular to $\widehat{(\z_0 - \bgamma_1(s))}$ \\
      $\Phi_{flat}^{WB}(\x)$ & Flattened wideband SAR interferometric phase \\
      $\phi(t)$ & Smooth windowing function \\
      $T_{\phi}$ & Duration of $\phi(t)$ \\
      $d_i^{UNB}(\mu,s)$ & Doppler-SAR data \\
      $\brmn L_i(\brmn z, s) \cdot \dot{\bgamma}_i(s) = \mathrm{C}$ & Iso-Doppler surface\\
      $\brmn L_i(\z,
      s)\cdot\ddot{\bgamma}_i(s)-\frac{\dot{\bgamma}_i(s)\cdot\dot{\bgamma}_i^\perp(s)}{R_i(\z, s)} = \mathrm{C}$ & Iso-Doppler-rate surface\\
      $\Kt_i^{UNB}$ & FBP operator for Doppler-SAR\\
      $I^{UNB}_i$ & Doppler-SAR image\\
      $\Phi^{UNB}_{s_d}(\brmn x)$ & Doppler-SAR interferometric phase\\
      $\brmn v$ & Baseline velocity \\
      $\Phi_{flat}^{UNB}(\x)$ & Flattened Doppler-SAR interferometric phase \\
      \hline \hline
    \end{tabular}}
\end{table}

\section{Wideband SAR Interferometry}
\label{sec:WidebandInterferometry}
The basic principles of SAR interferometry are described by many sources \cite{rosen2000synthetic}, \cite{zebker1994derivation},~\cite{hanssen2001radar}, \cite{madsen1993topographic} \cite{prati1990seismic}, \cite{bamler1998synthetic} and \cite{rodriguez1992theory}.  In this section, we summarize the principles and theory of SAR interferometry in a notation and context relevant to our subsequent presentation of Doppler-SAR interferometry. We begin with the wideband SAR received signal model, derive the interferometric phase model, provide a geometric interpretation of the interferometric phase from which we develop the equations of height mapping.

\subsection{Wideband SAR received signal model}
We assume that the SAR antennas are transmitting wideband waveforms.  Let $r_i(t,s)$ denote the received signals, $i=1,2$ where $s$ and $t$ are the \emph{slow-time} and \emph{fast-time} variables, respectively. Under the start-stop and Born approximations, the received signals can be modeled as \cite{NC03,yarman-hitchhiker,yarman-bistatic}:
\begin{eqnarray}
r_i(t,s)=\int \rme^{-\rmi\omega(t-2R_i(\x, s)/c)}\tilde A_i(\brmn x,s,\omega)V(\brmn x)\rmd\brmn x\rmd\omega
\label{eq:fwd1}
\end{eqnarray}
where
\begin{eqnarray}
R_i(\brmn x,s)=|\brmn x-\bgamma_i(s)|
\end{eqnarray}
is the range of the $i^{\textmd{th}}$ antenna, $c$ is the speed of light in free-space, $\omega$ is the temporal frequency variable, $V(\brmn x)$ is the scene reflectivity function. $\tilde A _i$ is a slowly-varying function of $\omega$ that depends on antenna beam patterns, geometrical spreading factors and transmitted waveforms.

Let $\omega=\omega_0+\omega'$, $\omega'\in\Omega$ where $\Omega$ is the bandwith and $\omega_0$ is the center frequency of the transmitted waveforms. We demodulate the received signals 
and write
\begin{eqnarray}
d_i^{WB}(t,s)&=&\rme^{\rmi\omega_0t}r_i(t,s),\\
&=&\int\limits_{-\Omega}^{\Omega} \rme^{-\rmi\omega'(t-2R_i(\brmn x,s)/c)}\tilde A _i(\brmn x,s,\omega')\rme^{\rmi2\frac{\omega_0}{c} R_i(\brmn x,s)}V(\brmn x)\rmd\brmn x\rmd\omega'.
\label{eq:fwd3}
\end{eqnarray}
Next, we approximate $R_i(\brmn x, s)$ in $\rme^{\rmi\frac{\omega_0}{c} R _i(\brmn x,s)}$
around $s=s_0^i$ as follows:

\begin{eqnarray}
R _i(\brmn x, s) \approx R _i(\brmn x, s_0^i)&+&(s-s_0^i)\left.\partial_s R _i(\brmn x,
  s)\right|_{s=s_0^i}\cr
  &+&\frac{(s-s_0^i)^2}{2}\left.\ \partial^2_s R _i(\brmn x, s) \right |_{s=s_0^i}, \ i=1,2
\end{eqnarray}
where $\partial_s$ denotes derivative with respect to $s$ and $s_0^i$ is the \emph{zero-Doppler time} for the $i^{\textmd{th}}$ antenna, i.e.,
\begin{equation}
\left.\partial_s R _i(\brmn x, s) \right |_{s=s_0^i} = \widehat{(\brmn x-\bgamma_i(s_0^i))}\cdot\dot{\bgamma}_i(s_0^i) = 0.
\label{eq:s_z}
\end{equation}
In (\ref{eq:s_z}) $\widehat{\x}$ denotes the unit vector in the direction of $\x$ and $\dot{\bgamma}_i(s)$ denotes the velocity of the $i^{\mathrm{th}}$ antenna.

  We define
\begin{equation}
\brmn L_i(\brmn x, s) = \widehat{(\brmn x-\bgamma _i(s))}
\label{eq:Look-dir}
\end{equation}
   and refer to $\brmn L_i(\brmn x, s)$ as the \emph{look-direction} of the $i^{\textmd{th}}$ antenna. Note that at the zero-Doppler time, the antenna look-direction is orthogonal to the antenna velocity.

Let
\begin{equation}
 A_i(\brmn x,s,\omega')=\tilde A _i(\brmn x,s,\omega')\rme^{\rmi 2\frac{\omega_0}{c}\frac{(s-s_0^i)^2}{2} \left [ \left.\partial^2_s R _i(s,\brmn x) \right|_{s=s_0^i}\right]}.
\end{equation}
Finally, we write the demodulated received signal as follows:
\begin{eqnarray}
d_i^{WB}(t,s)&\approx&\int\limits_{-\Omega}^{\Omega} \rme^{-\rmi\omega'(t-2R_i(\brmn x, s)/c)} A _i(\brmn x,s,\omega')\rme^{\rmi2\frac{\omega_0}{c} R _i(\brmn x, s_0^i)}V(\brmn x)\rmd\brmn x\rmd\omega'.
\label{eq:fwd4}
\end{eqnarray}

\subsection{Wideband SAR image formation and layover}
\begin{figure}[htpb]
\begin{center}
{\includegraphics[width=4in]
          {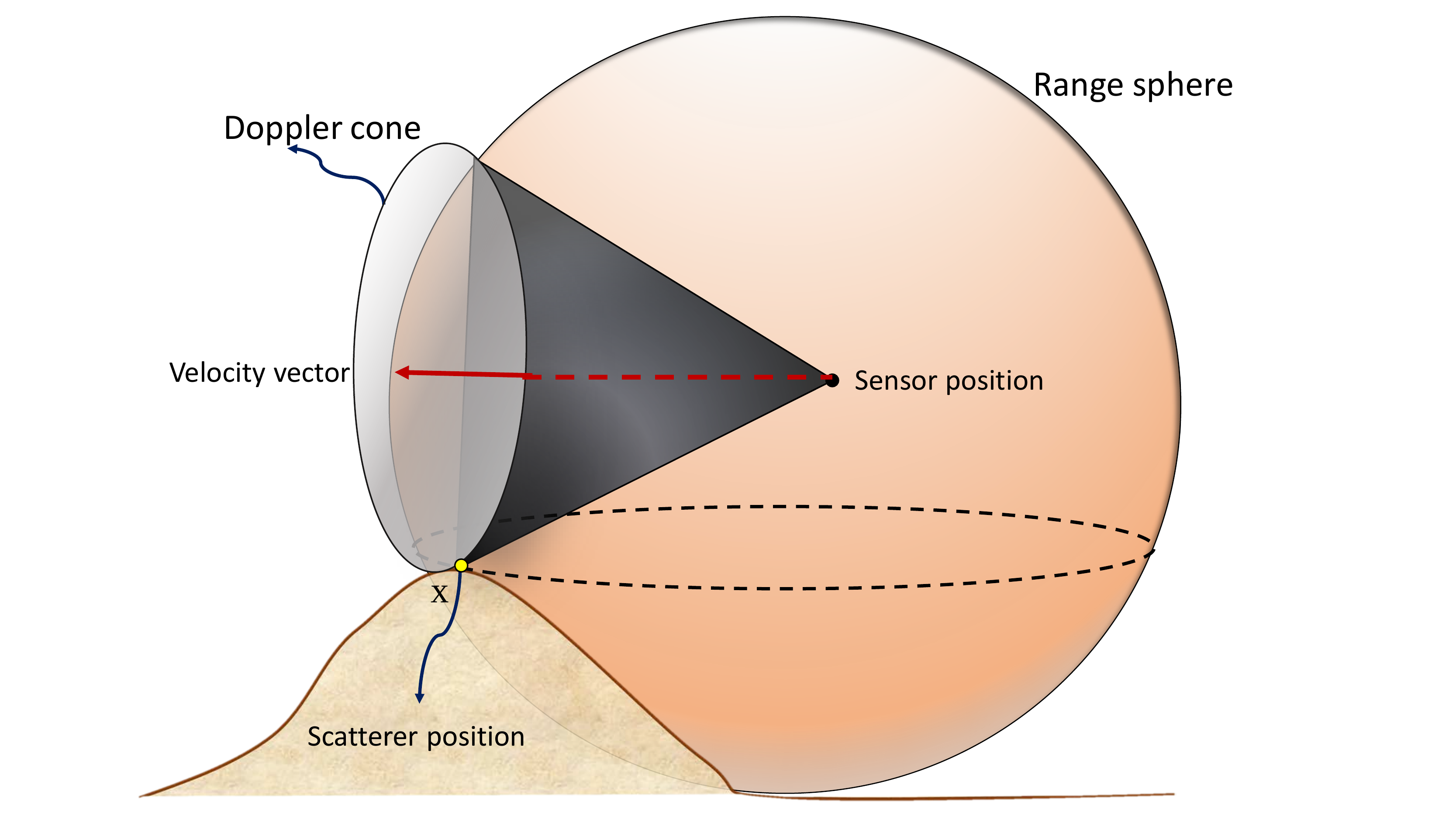}      }
\end{center}
 \caption{The SAR image of a scatterer is reconstructed at the intersection of the iso-range (sphere) and iso-Doppler (cone) surfaces and the height of the scatterer.}
\label{fig:Figure2}
\end{figure}

Many different algorithms were developed to form wideband SAR images such as range-Doppler \cite{zebker1994derivation}, seismic migration \cite{prati1990limits}, backprojection \cite{yarman-bistatic} and chirp scaling \cite{raney1994precision} algorithms. All of these algorithms take advantage of high range resolution provided by wideband transmitted waveforms and pulse-to-pulse Doppler information provided by the movement of antennas.  The location of a scatterer is identified by intersecting the iso-range and iso-Doppler surfaces and the ground topography as shown in Fig. \ref{fig:Figure2}.

More precisely, the image of a scatterer is formed at $\bi z$ satisfying the following equations:
\begin{eqnarray}
  \textmd{Iso-range} \ \textmd{surface:} \qquad\qquad\ \  |\z - \bgamma_i(s)| &= R_i(\x, s) \label{eq:InterIsoRangeDoppTopo1}\\
  \textmd{Iso-Doppler} \ \textmd{surface:} \quad \widehat{(\z-\bgamma_i(s))}\cdot\dot{\bgamma}_i(s) &= \partial_s R_i(\x, s) \label{eq:InterIsoRangeDoppTopo2}\\
  \textmd{Height:} \qquad\qquad\qquad\qquad\qquad\qquad\  z_3 &= h(\bi x), \ \ \z = [\bi z, z_3].
    \label{eq:InterIsoRangeDoppTopo}
\end{eqnarray}
Note that $R_i(\x, s)$ and $\partial_s R_i(\x, s)$ are the measured range and Doppler and $h(\bi x)$ is the height of the scatterer. As functions of $\z$, (\ref{eq:InterIsoRangeDoppTopo1}) and (\ref{eq:InterIsoRangeDoppTopo2}) define the iso-range and iso-Doppler surfaces, respectively.

 Iso-range contours are defined as the intersection of the iso-range surface, i.e., sphere, and the ground topography. Without loss of generality, we consider a filtered backprojection (FBP) type method where the received and demodulated signals are backprojected onto iso-range contours defined on a \emph{reference surface} \cite{yarman-bistatic}, \cite{NC03}. In the absence of heigh information, demodulated signal is backprojected onto the intersection of the iso-range surface and a known reference surface. Without loss of generality, we assume a flat reference surface at zero height and backproject the demodulated signals onto the following iso-range contours:
\begin{eqnarray}
H^{Range}_i(\z_0) = \left \{ \z_0 \in  \left. \FR^3 \ \right | \ \z_0 = [\bi z,\ 0] \ \textrm{and} \ |\z_0 - \bgamma_i(s)| = R_i(\brmn x, s)  \right \}.
\label{eq:Widebandiso-range}
\end{eqnarray}
Let $\Kt_i^{WB}$ be an FBP operator. Then, the reconstructed image of the scatterer at $\x$ becomes
\begin{eqnarray}
I_i^{WB}(\brmn z_{0}^i)&:=&\Kt_i^{WB}[\tilde{d}_i^{WB}](\brmn z_0^i),\cr
&=&\int\rme^{\rmi\omega'(t-2R _i(\z_{0}^i, s)/c)}Q_i^{WB}(\bi z_{0}^i,\omega', s){d}_i^{WB}(t,s)\rmd\omega'\rmd t\rmd s,
\label{eq:fbp}
\end{eqnarray}
where $Q_i^{WB}$ is a filter that can be chosen with respect to a variety of criteria \cite{yarman-bistatic}, \cite{yazici-cheney}.

\begin{figure}[htpb]
\begin{center}
{\includegraphics[width=3in]
          {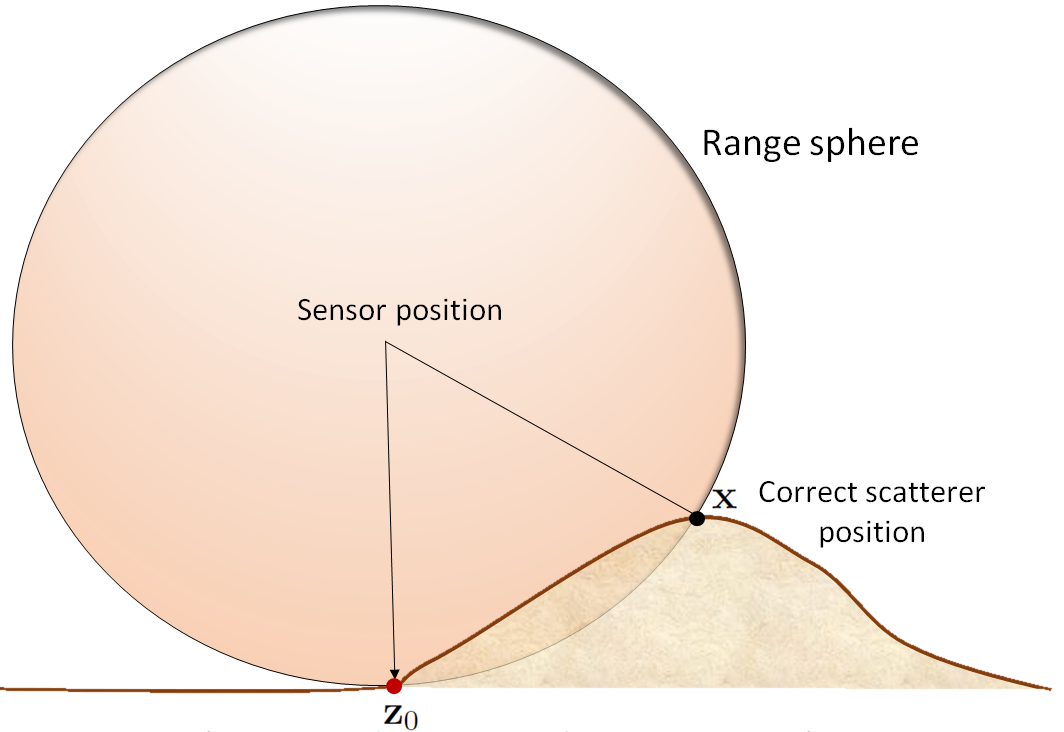}      }
\end{center}
 \caption{Layover in wideband SAR - The range sphere depicts the iso-range surface of the monostatic SAR configuration. Since the correct height of the scatterer at location $\x$ is unknown, the image of the scatterer at $\x$ is formed at $\z_0$ on a flat surface.}
\label{fig:lo-wb}
\end{figure}

From (\ref{eq:fwd4}), the image of the scatterer at $\x$ becomes
\begin{equation}
I_i^{WB}(\brmn z_{0}^i)=|I_i^{WB}(\brmn z_{0}^i)|\rme^{\rmi2\frac{\omega_0}{c} R _i(\brmn x,s_0^i)}.
\label{eq:dum}
\end{equation}
The magnitude of reconstructed images is a measure of target reflectivity, whereas the phase of the reconstructed image depends on the true location, $\x = [\bi x, \ h(\bi x)]$ of the scatterer. However, since the true height $h(\bi x)$ of the scatter is unknown and hence different than that of the reference surface, the location, $\bi z_0^i$, at which the scatterer is reconstructed is different than its true location, $\bi x$. This positioning error due to incorrect height information is known as \emph{layover}. Fig. \ref{fig:lo-wb} depicts the layover effect.

We see that without the knowledge of ground topography, additional information or measurements are needed to reconstruct the scatterers at correct locations. This additional information is provided by a second antenna that has a different vantage point than the first one.

\subsection{Wideband SAR interferometric height reconstruction}
An \emph{interferogram} is formed by multiplying one of the SAR images with the complex conjugate of the other SAR image \cite{hanssen2001radar,rosen2000synthetic}. Prior to multiplying the SAR images, the two intensity images, $|I_i^{WB}(\brmn z_0^i)|$, $i=1,2$ are co-registered so that pixel locations $\bi z_0^1$ and $\bi z_0^2$, each corresponding to the scatterer at position $\x$ in the scene, are roughly aligned\footnote{The positioning errors due to layover are different in the two SAR images due to different imaging geometries.}. Multiplying $I_1^{WB}(\brmn z_0^1)$ with the complex conjugate of $I_2^{WB}(\brmn z_0^2)$, we get
\begin{equation}\label{eq:SAR_WBint}
  I_1^{WB}(\brmn z_0^1) \overline{I_2^{WB}(\brmn z_0^2)} = |I_1^{WB}(\brmn z_0^1)||I_2^{WB}(\brmn z_0^2)| \mathrm{e}^{\mathrm{i}2\frac{\omega_0}{c}(R_1(\brmn x, s_0^1)-R_2(\brmn x, s_0^2))}.
\end{equation}
 We refer to the phase of the interferogram as the \emph{wideband interferometric phase}
\begin{eqnarray}\label{eq:hede1}
\Phi^{WB}_{s_0}(\x) &=& 2\frac{\omega_0}{c}(R_1(\brmn x, s_0^1)-R_2(\brmn x, s_0^2))
\end{eqnarray}
where $s_0$ is a multi-index for $\{s_0^1,s_0^2\}$.
The interferometric phase $\Phi^{WB}_{s_0}$ provides us the third measurement needed to determine the location of a scatterer in $\FR^3$. In general the range difference can be many multiples of $2\pi$. Unique phase proportional to range difference can be determined by a phase unwrapping process \cite{bamler1998synthetic}.

\begin{figure}[htpb]
\begin{center}
{\includegraphics[width=3in]
          {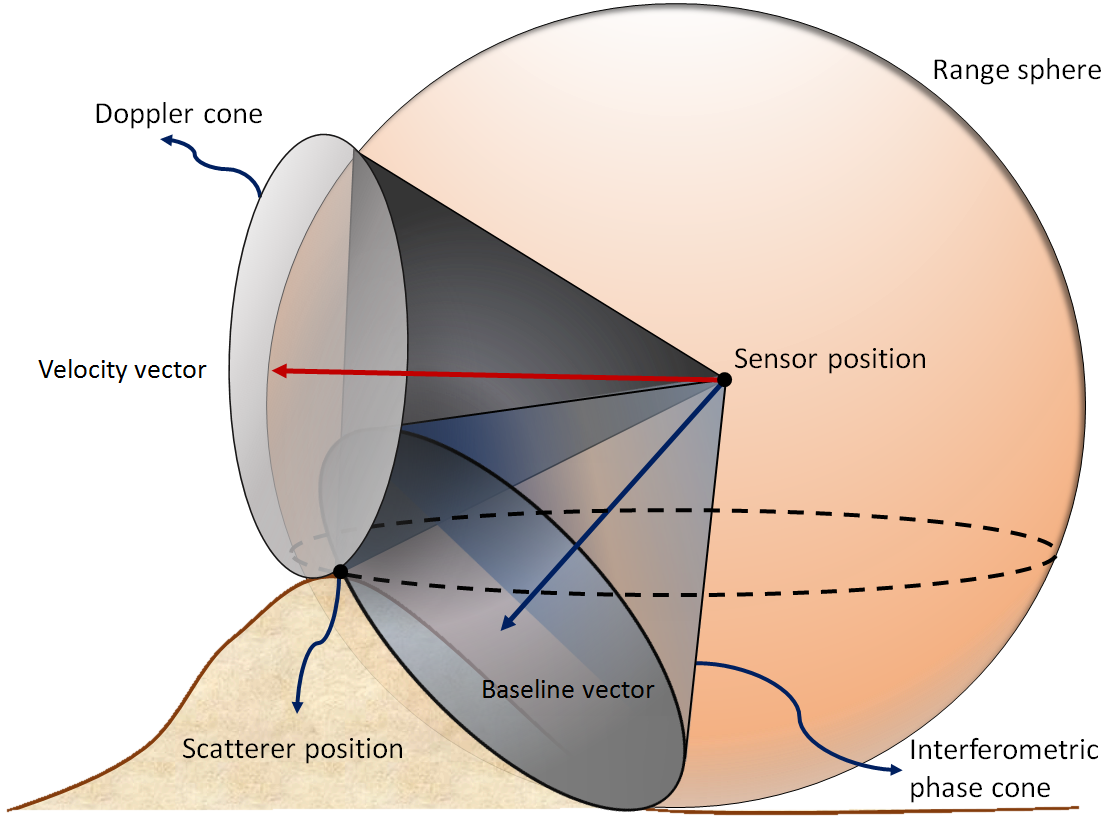}      }
\end{center}
 \caption{Wideband SAR interferometry provides a third algebraic equation by which the unknown location of a scatters in $\FR^3$ is determined. The scatterer is located at the intersection of the Doppler-cone, iso-range sphere, and the interferometric phase cone. The axis of rotation of the Doppler-cone is the velocity of the first antenna and the axis of rotation of the interferometric cone is the baseline vector extending from the first to the second antenna.}
\label{fig:wb-data}
\end{figure}

Now consider the following surface
\begin{eqnarray}
|\z - \bgamma_1(s_0^1)| - |\z - \bgamma_2(s_0^2)| = \frac{c}{2\omega_0}\Phi^{WB}_{s_0}(\x) \label{eq:hyperb}
\end{eqnarray}
where $\Phi^{WB}_{s_0}(\x)$ is the measured interferometric phase. (\ref{eq:hyperb}) defines a two-sheet hyperboloid with foci at $\bgamma_1(s_0^1)$ and $\bgamma_2(s_0^2)$. We assume that the distance between the antennas is much smaller than the ranges of the antennas to the scene and approximate this hyperboloid as follows:
\begin{eqnarray}
\brmn L_1(\brmn z, s_0^1)\cdot\brmn b \approx \frac{c}{2\omega_0}\Phi^{WB}_{s_0}(\x)
\label{eq:wb-int-phase}
\end{eqnarray}
where
\begin{equation}\label{eq:baseline}
  \brmn b=\bgamma_2(s_0^2)-\bgamma_1(s_0^1)
\end{equation}is the \emph{baseline vector}.
(\ref{eq:wb-int-phase}) defines a cone whose vertex is the first antenna and the axis of rotation is the baseline vector. We call this surface the \emph{interferometric phase cone}. The interferometric phase cone provides the third equation needed to locate the position of a scatterer in $\FR^3$. More precisely, the location of the scatterer is given by the solution of the following equations:
 \begin{eqnarray}
\textmd{Range} \ \textmd{sphere:}\qquad\qquad\qquad\ \  |\z - \bgamma_1(s_0^1)| &= R_1(\x, s_0^1) \label{eq:InterIsoRangeDoppTopo4}\\
\textmd{Doppler} \ \textmd{cone:}\qquad\ \ \   \widehat{(\z-\bgamma_1(s_0^1))}\cdot\dot{\bgamma}_1(s_0^1)) &= \left.\partial_s R_1(\x, s)\right|_{s=s_0^1} \label{eq:InterIsoRangeDoppTopo5}\\
\textmd{Interferometric} \ \textmd{phase} \ \textmd{cone:}\quad \brmn L_1(\brmn z, s_0^1)\cdot\brmn b  &= \frac{c}{2\omega_0}\Phi^{WB}_{s_0}(\x).
\label{eq:InterIsoRangeDoppTopo6}
\end{eqnarray}
The right-hand-side of (\ref{eq:InterIsoRangeDoppTopo4})-(\ref{eq:InterIsoRangeDoppTopo6}) are measured quantities defined in terms of the true location, $\x$, of the scatterer in the scene and the left hand-side-defines the three surfaces in terms of the location of the scatterer $\z$ in the image.
 Fig. \ref{fig:wb-data} geometrically illustrates the solution of these three equations in wideband SAR interferometry. 
Typically the variation in the color coding of interferogram is ``flattened'' by subtracting the expected phase from a surface of constant elevation. Let
$\brmn x = \brmn l+\brmn z_0$. Then, under the assumption that $|\l| \ll |\z_0 - \bgamma_1(s)|$
\begin{eqnarray}\label{eq:Appr_lookdirection}
  \widehat{(\x - \bgamma_1(s))} & \approx & \widehat{(\z_0 - \bgamma_1(s))} + \frac{\l_1^\bot}{|\z_0 - \bgamma_1(s)|}
\end{eqnarray}
where
\begin{equation}\label{eq:trasverse vec}
  \l_1^\bot = \l - \widehat{(\z_0 - \bgamma_1(s))}\left [\frac{\l \cdot \widehat{(\z_0 - \bgamma_1(s))}}{|\z_0 - \bgamma_1(s)|}\right].
\end{equation}In other words, the vector $\l_1^\bot$ is the component of $\l$ perpendicular to $\widehat{(\z_0 - \bgamma_1(s))}$. The flattened phase then becomes
\begin{eqnarray}
  \Phi_{flat}^{WB}(\x) &=& 2\frac{\omega_0}{c} \left[\L_1(\x, s_0^1) - \L_1(\z_0, s_0^1)\right] \cdot \b \\
   & \approx & 2\frac{\omega_0}{c} \frac{\l_1^\bot \cdot \b}{R(\z_0, s_0^1)}.
   \label{eq:flat_WB}
\end{eqnarray}
Since $\b_1^\perp \cdot \l = \l_1^\perp \cdot \b$ where  $\b_1^\perp$ is the component of $\b$ perpendicular to $\L_1(\z, s_0^1)$, (\ref{eq:flat_WB}) can be alternatively expressed as
\begin{equation}
\label{eq:flat_phase}
  \Phi_{flat}^{WB}(\x) = 2\frac{\omega_0}{c} \frac{\b_1^\bot \cdot \l}{R(\z_0, s_0^1)}.
\end{equation}

\begin{figure}[htpb]
\begin{center}
{\includegraphics[width=5in]
          {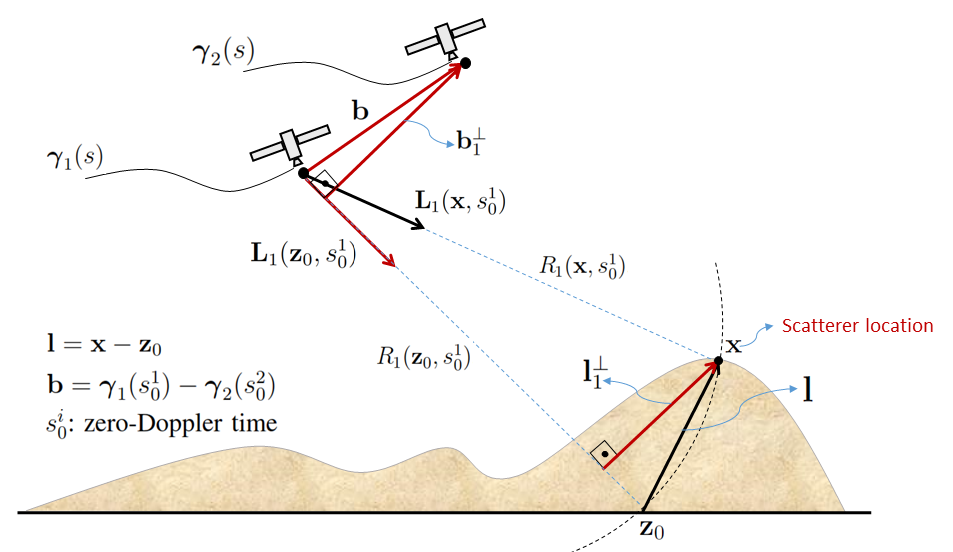}      }
\end{center}
 \caption{A two-dimensional illustration of vectors involved in wideband interferometric phase. $\brmn l=\brmn x-\brmn z_0$ where $\brmn z_0=[\bi z_0,0]$, $\bgamma_i(s_0^i)$ denotes the location of the $i^{th}$ antenna at the zero-Doppler time $s_0^i$, $\brmn L_1(\brmn z_0, s_0^1)$ denotes the look-direction of the first antenna with respect to the reference scatterer located at $\z_0$,  and $\brmn l^{\perp}_1$ is the component of  $\brmn l$ perpendicular to $\L_1(\brmn z_0, s_0^1)$. The wideband interferometric phase is related to the projection of the baseline vector, $\brmn b$, onto the the look-direction, $\brmn L_1(\brmn x, s_0^1)$, of the antenna with respect to scatterer location $\x$. Known vectors are shown in red and unknown vectors are shown in black.}
\label{fig:cs-wb}
\end{figure}

Fig. \ref{fig:cs-wb} illustrates the key concepts and vectors involved in the wideband interferometry.
%

\section{Data Model and Image Formation for Doppler-SAR}
\label{sec:Doppler-SAR}
\subsection{Data Model for Doppler-SAR}
We consider two mono-static antennas following the trajectories  $\bgamma_i(t)$, $i=1,2$, transmitting ultra-narrowband CWs as shown in Fig. \ref{fig:wb-geo}. 
Let $p(t)\approx\tilde p(t)\rme^{\rmi\omega_0}$ be the transmitted waveform where $\omega_0$ is the center frequency.
The scattered field model at the $i^\textmd{th}$ antenna is then given by
\begin{eqnarray} \label{eq:r_i}
r_i(t)=\frac{\omega_0^2}{(4\pi)^2}\int\frac{\rme^{-\rmi\omega_0(t- 2|\x - \bgamma_i(t)|/c)}}{|\brmn x-\bgamma_i(t)|^2}\tilde p(t-2|\x - \bgamma_i(t)|/c) V (\brmn x)\rmd \brmn x.
\end{eqnarray}
Let $\mu\in\FR^+$ and $\phi(t)$ be a smooth windowing function with a finite support,  $t\in[0,T_\phi]$. Following \cite{wang2013ground, wang2012bistatic, borden2005synthetic}, we correlate $r_i(t)$ with a scaled and translated version of the transmitted signal over $\phi(t)$ as follows:
\begin{eqnarray}
\label{eq:data}
d_i^{UNB}(\mu,s)=\int r_i(t)\rme^{\rmi\omega_0\mu(t-sT_\phi)/c}\tilde p^*(\mu (t-sT_\phi))\phi(t-sT_\phi)\rmd t.
\end{eqnarray}
Inserting (\ref{eq:r_i}) into (\ref{eq:data}), we obtain
\begin{eqnarray} \label{eq:d_i^UNB}
d_i^{UNB}(\mu,s)=\frac{\omega_0^2}{(4\pi)^2}& \int\frac{\rme^{-\rmi\omega_0(t- 2|\x - \bgamma_i(t)|/c)}}{|\brmn x-\bgamma_i(t)|^2}\tilde p(t-2|\x - \bgamma_i(t)|/c) V (\brmn x)\cr
& \quad \times\rme^{\rmi\omega_0\mu(t-sT_\phi)/c}\tilde p^*(\mu (t-sT_\phi))\phi(t-sT_\phi)\rmd t\rmd \brmn x.
\end{eqnarray}
Approximating $\bgamma_i (t)$ around $t=sT_\phi$, $\bgamma _i(t)\approx\bgamma _i(sT_\phi)+\dot{\bgamma}_i(sT_\phi)(t-sT_\phi)$, and making the far-field approximation, we write
\begin{eqnarray} \label{eq:farfiled4}
|\brmn x-\bgamma_i(t)|\approx |\brmn x-\bgamma_i(sT_\phi)|- \brmn L_i(\brmn x,sT_\phi)\cdot\dot{\bgamma}_i(sT_\phi)(t-sT_\phi),
\end{eqnarray}
where $\brmn L_i(\brmn x,sT_\phi)=\widehat{(\brmn x-\bgamma_i(sT_\phi))}$ and $\dot{\bgamma}_i(sT_\phi) = \partial_s\bgamma_i(sT_\phi)$ is the velocity of the $i^{\textmd{th}}$ antenna.

To simplify our notation, for the rest of the paper, we set $\brmn L _i(\brmn x,sT_\phi) = \brmn L_i(\brmn x,s)$, $\bgamma_i(sT_\phi) = \bgamma_i(s)$, $\dot{\bgamma}_i(sT_\phi) = \dot{\bgamma}_i(s)$, $\partial_s^2\bgamma_i(sT_\phi) = \ddot{\bgamma}_i(sT_\phi) = \ddot{\bgamma}_i(s)$ and $R_i(\x, sT_\phi) = R_i(\x,s)$.
We next define Doppler for the $i^{\textmd{th}}$ antenna
\begin{equation} \label{eq:DefinitionDopp}
f_i^d(\brmn x,s)= -\frac{\omega_0}{c}\brmn L_i(\brmn x,s)\cdot\dot{\bgamma}_i(s).
\end{equation}
Inserting (\ref{eq:farfiled4}) and (\ref{eq:d_i^UNB}) into (\ref{eq:DefinitionDopp}), the data model becomes
\begin{eqnarray}
d_i^{UNB}(\mu,s)=\int\rme^{-\rmi t[\omega_0(1-\mu)-2f_i^d(\brmn x,s)]}\tilde A _i(t,\brmn x,s,\mu)\rme^{\rmi 2f_i^d(\brmn x,s)sT_\phi} V (\brmn x)\rmd t\rmd \brmn x,
\end{eqnarray}
where $\tilde A _i(t,\brmn x,s,\mu)$ is a slow varying function of $t$ composed of the rest of the terms in (\ref{eq:d_i^UNB}).

We now approximate $f_i^d(\brmn x,s)$ around $s=s_d^i$ as follows:
\begin{equation} \label{eq:Zero-Dopp-rateApp}
f_i^d(\brmn x,s) \approx  f_i^d(\brmn x,s_d^i)+(s-s_d^i)\left.\partial_s f_i^d(\brmn x,s)\right|_{s=s_d^i} + \frac{(s-s_d^i)^2}{2}\left. \partial^2_s f_i^d(\brmn x,s)\right|_{s=s_d^i}.
\end{equation}
We choose $s_d^i$ such that
\begin{equation}
\left.\frac{\partial f_i^d(\brmn x,s)}{\partial s}\right|_{s=s_d^i} = 0 \quad \Rightarrow \quad
\brmn L_i(\brmn x, s_d^i)\cdot\ddot{\bgamma}_i(s_d^i)-\frac{\dot{\bgamma}_i(s_d^i)\cdot\dot{\bgamma}_i^\perp(s_d^i)}{R_i(\x s_d^i)}=0
\label{eq:s_z_dop}
\end{equation}
where $\ddot{\bgamma}_i(s_d^i)$ is the acceleration of the $i^{\textmd{th}}$ antenna and $\dot{\bgamma}_i^\perp(s_d^i)$ is the component of $\dot{\bgamma}_i(s_d^i)$ perpendicular to the look-direction $\brmn L_i(\brmn x,s_d^i)$ as described in (\ref{eq:trasverse vec}). We refer to $s_d^i$ as the \emph{zero-Doppler-rate time} for the $i^{\textmd{th}}$ antenna.

Using (\ref{eq:Zero-Dopp-rateApp}) in $\rme^{\rmi 2f_i^d(\brmn x,s)sT_\phi}$ and redefining the slow-varying function in $t$,
\begin{equation}
A _i(t,\brmn x,s,\mu)=\tilde A _i(t,\brmn x,s,\mu)\rme^{\rmi 2s_d^i T_\phi
\frac{(s-s_d^i)^2}{2}\partial^2_s f_i^d(\brmn x,s)|_{s=s_d^i}},
\end{equation}
we obtain the following data model for Doppler-SAR image reconstruction:
\begin{eqnarray} \label{eq:finalDoppforward}
d_i^{UNB}(\mu, s) \approx \int\rme^{-\rmi t[\omega_0(1-\mu)-2f_i^d(\brmn x,s)]} A_i(t,\brmn x,s,\mu)\rme^{\rmi 2f_i^d(\brmn x,s_d^i)s_d^iT_\phi} V (\brmn x)\rmd t\rmd \brmn x.
\end{eqnarray}


\subsection{Doppler-SAR Image Formation and Layover}
Similar to the wideband case, we reconstruct images by backprojection as described in \cite{wang2011doppler,wang2013ground} \cite{wang2012bistatic}. The forward model in (\ref{eq:finalDoppforward}) shows that the data, $d_i^{UNB}(s,\mu)$, is the weighted integral of the scene reflectivity over iso-Doppler contours.  It was shown in \cite{wang2012bistatic} that a scatterer located at $\x$ in the scene is reconstructed at the intersection of iso-Doppler surface and iso-Doppler-rate surface and ground topography.  More precisely, the image of a scatterer located at $\x$ in the scene is reconstructed at $\bi z$ satisfying the following equations:
 \begin{eqnarray}
\fl\qquad\textmd{Iso-Doppler} \ \textmd{surface:}\qquad\qquad\qquad\qquad\ \ \ \  \L_i^d(\z, s) \cdot \dot{\bgamma}_i(s) &= \frac{c}{\omega_0}f_i^d(\x, s) \label{eq:InterIsoDoppTopo1}\\
\fl\qquad\textmd{Iso-Doppler-rate} \ \textmd{surface:}\quad \L_i(\z, s)\cdot\ddot{\bgamma}_i(s)-\frac{\dot{\bgamma}_i(s)\cdot\dot{\bgamma}_i^\perp(s)}{R_i(\z, s)} &= \frac{c}{\omega_0} \partial_s f_i^d(\x, s) \label{eq:InterIsoDoppTopo2}\\
\fl\qquad\textmd{Height:}\qquad\qquad\qquad\qquad\qquad\qquad\qquad\qquad\qquad\ \ \ \  z_3 &= h(\bi x), \ \ \z = [\bi z, z_3]
\label{eq:InterIsoDoppTopo}
\end{eqnarray}
where the right-hand-side of (\ref{eq:InterIsoDoppTopo1})-(\ref{eq:InterIsoDoppTopo2}) corresponds to measurements and the left-hand-side defines surfaces in image parameter $\z$.

\begin{figure}[htpb]
\begin{center}
{\includegraphics[width=3.5in]
          {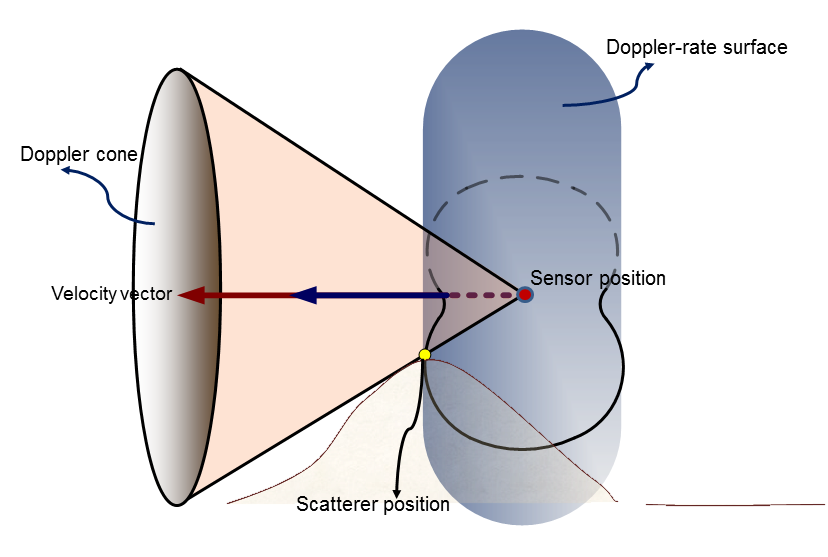}}
\end{center}
 \caption{In Doppler-SAR image reconstruction, a scatterer located at $\x$ in the scene is correctly reconstructed at the intersection of the iso-Doppler and iso-Doppler-rate surfaces and the ground topography. Iso-Doppler surface is a cone in which its vertex is the antenna location and its axis of rotation is the antenna velocity. The geometry of the iso-Doppler-rate surface depends on the antenna trajectory. Figure is drawn for a linear trajectory at a constant height. }
\label{fig:UNB_Recon}
\end{figure}

The iso-Doppler-rate surface, given by the following set,
 \begin{equation}
\fl\qquad H^{Dop-rate}_i(\z) = \left \{ \z \in  \left. \FR^3 \ \right | \ \brmn L_i(\z, s)\cdot\ddot{\bgamma}_i(s)-\frac{\dot{\bgamma}_i(s)\cdot\dot{\bgamma}_i^\perp(s)}{R_i(\z, s)} = \frac{c}{\omega_0}\partial_s f_i^d(\x, s) \right \}.
\label{eq:UNBiso-Dopp-rate}
\end{equation}
 can be viewed as a continuum of intersections of cones and expanding spheres centered at the sensor location. The axis of rotation for the surface is the acceleration vector of the antenna trajectory. Fig. \ref{fig:UNB_Recon} illustrates iso-Doppler and iso-Doppler-rate surfaces and the reconstruction of a point scatterer by the intersection of these surfaces and ground topography. The reconstruction is analogous to the wideband SAR image reconstruction shown  in Fig. \ref{fig:Figure2}.

In the absence of ground topography information, we backproject data onto iso-Doppler contours on a reference surface. Without loss of generality, we consider the following iso-Doppler contours:
 \begin{equation}
\fl\qquad\qquad H^{Dop}_i(\z_0) = \left \{ \z_0 \in  \left. \FR^3 \ \right | \ \z_0 = [\bi z,\ 0] \ \textrm{and} \ \brmn L_i(\brmn z,s)\cdot\dot{\bgamma}_i(s) = \frac{c}{\omega_0}f_i^d(\brmn x,s)  \right \}
\label{eq:UNBiso-Dopp}
\end{equation}
where the right-hand-side of the equality in (\ref{eq:UNBiso-Dopp}) is the high resolution measurement provided by ultra-narrowband CW.

\begin{figure}[htpb]
\begin{center}
{\includegraphics[width=3.0in]
          {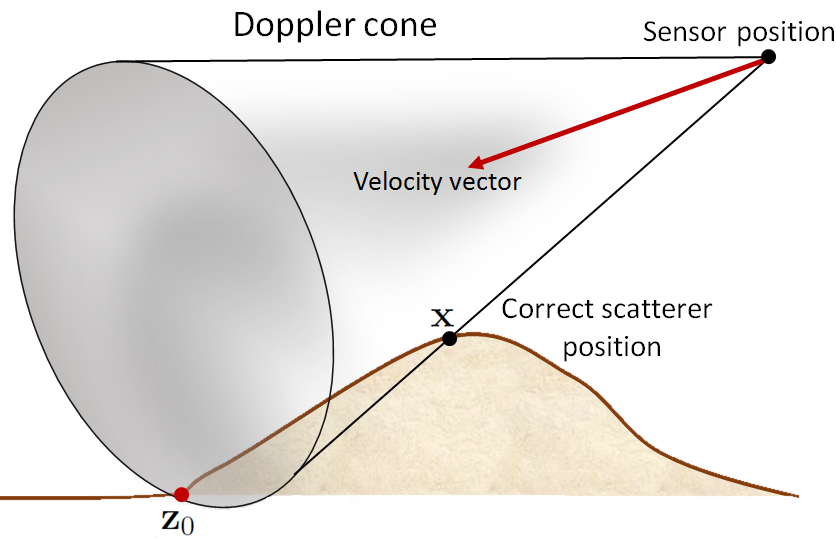}      }
\end{center}
 \caption{If the height of a scatter is not known, it is reconstructed at an incorrect position. Both the correct scatterer location $\x$ and its image $\z_0$ lie on the same iso-Doppler surface, i.e., the Doppler cone. $\brmn z_0$ lies at the intersection of the Doppler cone defined $f_1^d(\brmn x, s)$ and the flat topography.}
\label{fig:lo-dsar}
\end{figure}

Let  $\Kt_i^{UNB}$ be an FBP operator as described in \cite{wang2012bistatic}. Then, the reconstructed image is given by:
\begin{eqnarray}
I^{UNB}_i(\brmn z_0^i)&:= &\Kt_i^{UNB}[d^{UNB}_i](\brmn z_0^i)\cr
&\approx&\int\rme^{\rmi t(\omega_0(1-\mu)-2f_i^d(\z_0^i,s))}Q^{UNB}_i(s,\bi z_0^i,t)d^{UNB}_i(s,\mu) \rmd t\rmd\mu \rmd s
\label{eq:doppler s1}
\end{eqnarray}
where $Q^{UNB}_i$ is a filter that can be chosen as in \cite{wang2011doppler,wang2013ground,wang2012bistatic}. The reconstructed image is given by
\begin{equation}
I_i^{UNB}(\brmn z_0^i)=|I_i^{UNB}(\brmn z_0^i)|\rme^{\rmi 2 f_i^d(\brmn x,s_d^i)s_d^i T_\phi}.
\label{eq:dum2}
\end{equation}
In the absence of topography information, we see that a scatterer located at $\x$ in the scene is reconstructed at $\z_0^i \neq \x$ in the image. This position error in the reconstructed image is the counterpart of the layover effect observed in conventional wideband SAR images. Fig. \ref{fig:lo-dsar} illustrates the layover effect in Doppler-SAR. However, the phase of the reconstructed image is a function of the scatterer's true location, $\x$, and hence, includes its height information, $h(\bi x)$.

Note that the phases of the reconstructed images depend on the Doppler-rate, $f_i^d(\brmn x,s_d^i)$, the duration of the windowing function, $T_{\phi}$, and the corresponding zero-Doppler-rate times, $s_d^i$. The height information is included in the Doppler-rate. However, since each imaging geometry may yield different zero-Doppler-times, Doppler-rate in the phase of each image is multiplied by a different zero-Doppler-rate time. To equalize the effect of this multiplication factor, we multiply one of the reconstructed images with itself so that the Doppler-rate in the phase of both images are multiplied by the same factor, say $s_d^1$. As a result, each image becomes
\begin{equation}
I_i^{UNB}(\brmn z_0^i)=|I_i^{UNB}(\brmn z_0^i)|\rme^{\rmi 2 f_i^d(\brmn x,s_d^i)s_d^1 T_\phi}, \quad i=1,2.
\label{eq:dum3}
\end{equation}

\section{Doppler-SAR Interferometric Height Reconstruction}
\label{sec:Doppler-SAR_interferometry}
Similar to the wideband case, we form two Doppler-SAR images, $I_i^{UNB}(\brmn z_{0}^i)$, $i=1,2$, co-register the intensity images $|I_i^{WB}(\brmn z_{0}^i)|$ 
and multiply one of them by the complex conjugate of the other to form an interferogram. Then the interferometric phase, i.e., the phase function of  $I_1^{UNB}(\brmn x)\overline{I_2^{UNB}(\brmn x)}$ is given by
\begin{equation}\label{eq:DoppInterPhase}
  \Phi^{UNB}_{s_d}(\brmn x) = 2s_d^1T_\phi\left(
f_1^d(\brmn x,s_d^1)-f_2^d(\brmn x,s_d^2)\right)
\end{equation}
where $s_d$ denotes multi-index for $\{s_d^1, s_d^2\}$. Thus the scatterer lies on the following surface:
\begin{equation}\label{eq:DoppInterPhase_surf}
  \L_1(\z,s_d^1)\cdot \dot{\bgamma}_1(s_d^1) - \L_2(\z,s_d^2)\cdot \dot{\bgamma}_2(s_d^2) = -\frac{c}{2\omega_0}\left (f_1^d(\brmn x,s_d^1)-f_2^d(\brmn x,s_d^2) \right)
\end{equation}
where the right-hand-side is the measured interferometric phase. The left-hand-side of (\ref{eq:DoppInterPhase_surf}) defines a surface that can be described as the intersections of two cones one of which has a continuously changing solid angle.

 Assuming that the distance between the antennas is much smaller than the ranges of the antennas to the scene, we can approximate the look-direction of the second antenna in terms of the look-direction of the first one as follows:
\begin{equation}\label{eq:LookDirApp}
  \L_2(\x, s_d^2) = \L_1(\x, s_d^1) + \frac{\b_1^\perp}{R_1(\x, s_d^1)}
\end{equation}
where $\b = \bgamma_2(s_d^2) - \bgamma_1(s_d^1)$ is the baseline vector and $\b_1^\perp$ is the component of $\b$ perpendicular to the look-direction of the first antenna.
Using (\ref{eq:LookDirApp}), we approximate the interferometric phase as follows:
\begin{equation}\label{eq:DoppInterPhase2}
  -\frac{c}{2s_d^1T_\phi\omega_0}\Phi^{UNB}_{s_d}(\brmn x) \approx \L_1(\x,s_d^1)\cdot \v + \frac{\b_1^\perp \cdot \dot{\bgamma}_2(s_d^2)}{R_1(\x,s_d^1)}
\end{equation}
where
\begin{equation}\label{eq:baselineVelocity}
  \v = \dot{\bgamma}_2(s_d^2) - \dot{\bgamma}_1(s_d^1).
\end{equation} We refer to $\v$ as the \emph{baseline velocity}.  We see that (\ref{eq:DoppInterPhase2}) approximates the interferometric phase as a Doppler-rate. Additionally, (\ref{eq:DoppInterPhase2}) shows that Doppler-SAR interferometry involves not only configuring antennas in position space, but also in velocity space. The larger the difference in antenna velocities in the look-direction of the first antenna, the larger the interferometric phase becomes.  If on the other hand, the velocities of the antennas are the same, the second term in (\ref{eq:DoppInterPhase2}) defines the interferometric phase surface.

\begin{figure}[htpb]
\begin{center}
{\includegraphics[width=3.5in]
          {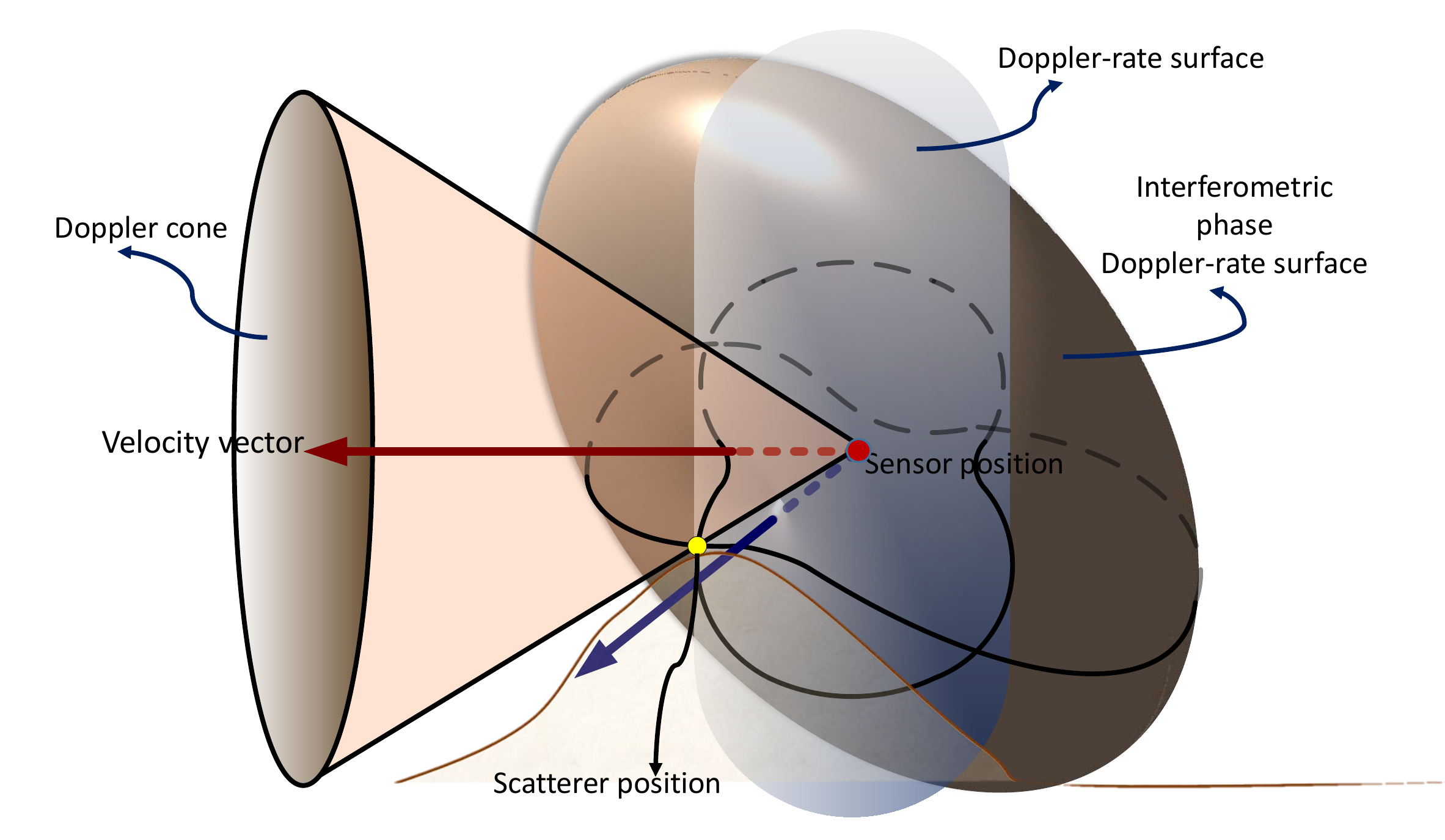}}
\end{center}
 \caption{Determination of the scatterer location in Doppler-SAR interferometry. The scatterer is located at the intersection of the Doppler cone and the two iso-Doppler-rate surfaces. Interferometric phase measurement provides the third surface, i.e., the interferometric phase iso-Doppler-rate surface. }
\label{fig:dsar-int}
\end{figure}

Clearly, in Doppler-SAR interferometry (\ref{eq:DoppInterPhase2}) provides the third equation needed to determine the location of a scatterer in $\FR^3$. More precisely, the location of a scatterer is given by the solution of the following three equations:

 \begin{eqnarray}
\fl\textmd{Iso-Doppler:} \qquad\qquad\qquad\qquad\qquad\   \widehat{(\z - \bgamma_1(s_d^1))}\cdot\dot{\bgamma}_1(s_d^1) &= \frac{c}{\omega_0}f_1^d(\x, s_d^1) \label{eq:InterIsoDoppTopo3}\\
\fl\textmd{Iso-Doppler-rate:}\ \widehat{(\z - \bgamma_1(s_d^1))}\cdot\ddot{\bgamma}_1(s_d^1)-\frac{\dot{\bgamma}_1(s_d^1)\cdot\dot{\bgamma}_1^\perp(s_d^1T_\phi)}{R_1(s_d^1,\z)} &= \partial_s f_1^d(\x, s_d^1) \label{eq:InterIsoDoppTopo4}\\
\fl \textmd{Interferometric Doppler-rate:}\quad\  \L_1(\z,s_d^1)\cdot \v + \frac{\b_1^\perp \cdot \dot{\bgamma}_2(s_d^2)}{R_1(\z,s_d^1)}  &= -\frac{c}{2s_d^1T_\phi\omega_0}\Phi^{UNB}_{s_d}(\x).
\label{eq:InterIsoDoppTopo5}
\end{eqnarray}
Fig. (\ref{fig:dsar-int}) depicts the intersection of the three surfaces at the scatterer location in $\FR^3$.

Similar to the wideband SAR interferometry, the interferometric phase can be ``flattened'' by subtracting the phase due to a scatterer with known height. Without loss of generality, let $\z_0 = [\bi z, 0]$ with $R_1(\z_0, s) = R_1(\x, s)$ and $\x = \z_0 + \l$. Thus, identifying the location of a scatterer is equivalent to determining $\l$.

\begin{figure}[htpb]
\begin{center}
{\includegraphics[width=5in]
          {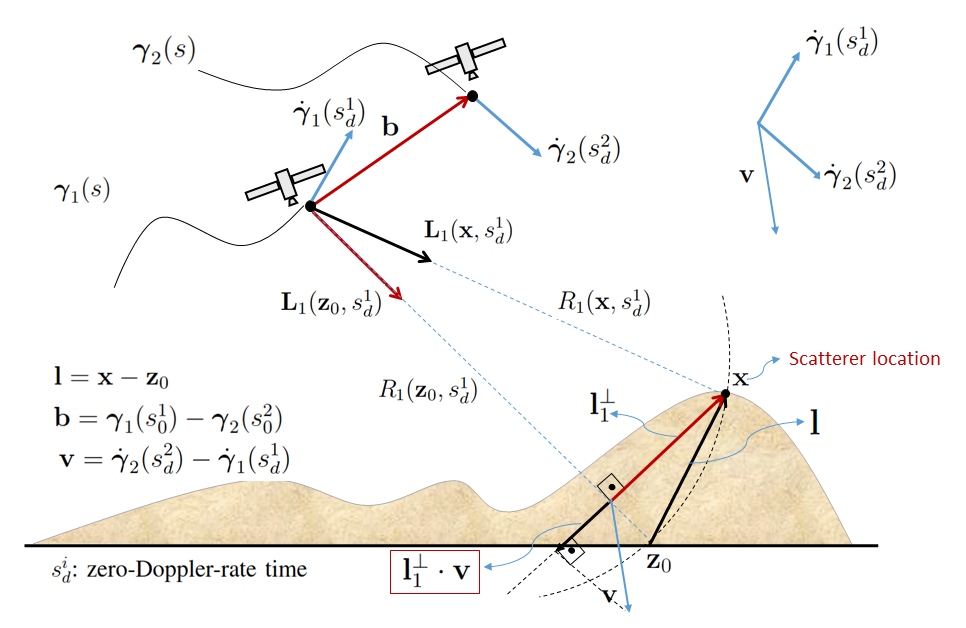}      }
\end{center}
 \caption{An illustration of key concepts and vectors in Doppler-SAR interferometry. $\brmn l=\brmn x-\brmn z_0$ where $\brmn z_0=[\bi z_0,0]$, $\bgamma_i(s)$ denotes the $i^{\textmd{th}}$ antenna position, $\brmn L_1(\brmn z_0, s_d^1)$ denotes the look-direction with respect to a reference surface,  $\brmn l^{\perp}_1$ is the component of $\brmn l$ perpendicular to $\brmn L_1(\brmn z_0, s_d^1)$, $\dot{\bgamma}_1(s)$ denotes the antenna velocity, $\brmn L_1(\brmn x, s_d^1)$ denotes the look-direction of the antenna with respect to the correct target location. Doppler-SAR interferometric phase is proportional to the projection of the baseline velocity vector onto $\brmn l^{\perp}_1$. Known vectors are shown in red and unknown vectors are shown in black.}
\label{fig:cs-dsar}
\end{figure}

Using (\ref{eq:Appr_lookdirection}), we see that
\begin{equation}\label{eq:Phi_UNB_flat}
  \Phi_{flat}^{UNB}(\x) = \Phi_{s_d}^{UNB}(\x) - \Phi_{s_d}^{UNB}(\z_0) \approx \frac{\l_1^\perp \cdot \v}{R_1(\z_0, s_d^1)} + \mathcal{O}\left(\frac{1}{R_1^2(\z_0,s_d^1)}\right)
\end{equation}
where $\l_1^\perp$ is the component of $\l$ perpendicular to $\L_1(\z_0, s_d^1)$.
\eqref{eq:Phi_UNB_flat} shows that the flattened interferometric phase for Doppler-SAR interferometry is related to the projection of the unknown $\l_1^\perp$ onto the baseline velocity vector scaled by the range of the first antenna to $\z_0$. Since $ \l_1^\perp \cdot \v = v_1^\perp \cdot \l$ where $v_1^\perp$ is the component of $\v$ perpendicular to $\L_1(\z_0, s_d^1)$, we alternative express (\ref{eq:Phi_UNB_flat}) as follows:
\begin{equation}\label{eq:Phi_UNB_flat2}
  \Phi_{flat}^{UNB}(\x) \approx \frac{\v_1^\perp \cdot \l}{R_1(\z_0, s_d^1)}.
\end{equation}
Fig. \ref{fig:cs-dsar} shows the key concepts and vectors involved in Doppler-SAR interferometry.

\subsection{Comparison of Doppler-SAR Interferometry wide Wideband Case}
Table II tabulates the interferometric phase for the wideband SAR and Doppler-SAR cases.  We
compare and contrast the two interferometric phases below:
\begin{itemize}
\item For WB and UNB, the ``baseline'' is the difference in range and difference in velocity,
  respectively.
\item The larger the $\omega_0$, the center frequency, the larger the interferometric phase in both
  WB and UNB cases.
\item The larger the range, $R_1$, the smaller the interferometric phase in both WB and UNB cases.
\item For UNB, larger the $T_{\phi}$, the larger the interferometric phase.
\item For WB, the larger the $\b$, the difference between the positions of the two antennas, the
  larger the interferometric phase.  For UNB, the larger the $\v$, the difference between the velocities of the two antennas, the
  larger the interferometric phase.
\end{itemize}

\begin{table}[htp] \label{Table:Comp}
\caption{Raw and flattened interferometric phase functions for wideband SAR and Doppler-SAR.}
\begin{center}
{\footnotesize
    \begin{tabular}{ l | p{7cm} | p{5cm}| }
\cline{2-3}
         & Interferometric Phase & Flattened Interferometric Phase \\ \hline \hline
\multicolumn{1}{ |l| }{Wideband SAR} & $2\frac{\omega_0}{c} \left [\L_1(\x, s_0^1)\cdot\b \right ]$ &
$2\frac{\omega_0}{c}\frac{1}{R_1(\z_0, s_0^1)} \left [\brmn b^{\perp}_1 \cdot \l \right ]$\\ \hline
\multicolumn{1}{ |l| }{Doppler-SAR} & $
-2\frac{\omega_0}{c}s_dT_\phi \left [\L_1(\x, s_d^1)\cdot \v + \frac{1}{R_1(\x, s_d^1)} \b_1^\perp \cdot \dot{\bgamma}_2(s_d^2) \right ]$ & $-2\frac{\omega_0}{c}s_dT_\phi \frac{1}{R_1(\z_0, s_d^1)} \left [\v_1^\perp \cdot \l \right ] $\\
        \hline
    \end{tabular}}
\end{center}
\label{tab:int_phase}
\end{table}


\section{Numerical Experiments}\label{sec:simulations}
\subsection{Experimental Setup}
We conducted numerical experiments for both wideband and Doppler-SAR.  Our experimental setup was
as follows:
\begin{itemize}
\item A scene of size $128\times 128$m at $1$m resolution was imaged.
\item A single point target was placed at $(-20,-31,50)$m with the origin $(0,0,0)$ at the scene
  center.
\item Two antennas flying on a linear trajectory parallel to the $y$-axis was used with both
  antennas placed at $7.1$km from the scene center in the $x$-axis direction.  The midpoint of the
  linear trajectories for both antennas was aligned at $y=0$.
\item \textbf{Wideband:}
  First antenna was placed at height of $3$km and the second at $4$km.  The length of the
    trajectories were $1$km in length for both antennas.
  Both antennas were moving at velocity of $100$m/s.
  A waveform with flat spectrum of 100$MHz$ bandwidth at center frequency of $8$GHz was
    transmitted from both antennas.
  $512$ frequency samples and $1024$ slow-time, $s$, samples were used for imaging.
\item \textbf{Doppler:}
  First antenna was placed at height of $2$km and the second at $4$km.  The length of the
    trajectories were $1$km for both antennas.
  The first antenna was moving at velocity of $100$m/s and the second at $400$m/s.
  A continuous waveform at center frequency of $8$GHz was transmitted from both antennas.
  A window of $0.01$s was used for processing at each slow time.
  $512$ fast time, $t$, samples and $1024$ slow-time, $s$, samples were used for imaging.
\end{itemize}

\subsection{Wideband SAR Interferometry}
\begin{figure}[htpb]
  \subfloat[][]{
    \includegraphics[width=0.52\textwidth]
    {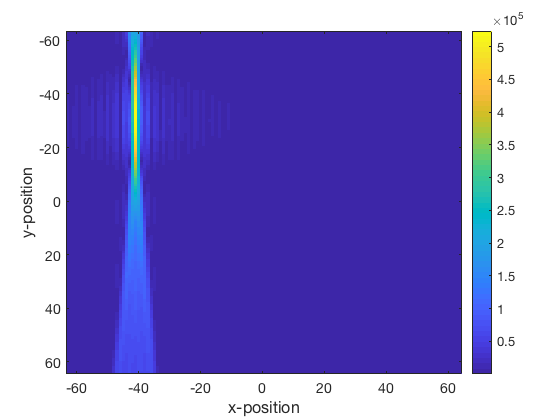}\label{fig:scene1_wb}}
  \subfloat[][]{
    \includegraphics[width=0.52\textwidth]
          {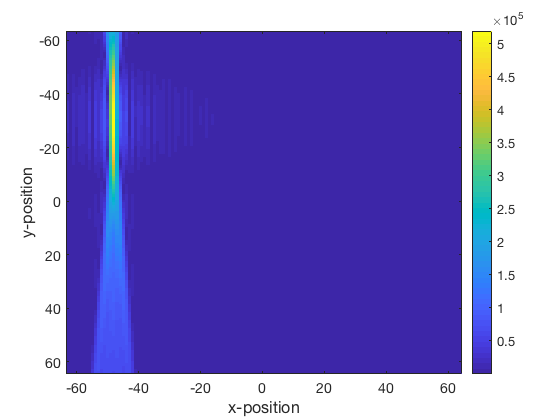}\label{fig:scene2_wb}}
        \caption{(a) Wideband reconstruction of the target located at $(-20,-31,50)$m using the first antenna assuming flat ground topography.  The target is
  reconstructed at $(-41,-31,0)$m.  (b) Wideband reconstruction of the target located at $(-20,-31,50)$m using the second antenna assuming flat ground topography.  The target is
  reconstructed at $(-48,-31,0)$m.}
\end{figure}

Fig.~\ref{fig:scene1_wb} and Fig.~\ref{fig:scene2_wb} show the reconstructed images of the point
target located at $(-20,-31,50)$m from the first and the second antenna, respectively assuming a
flat ground topography at height of $0$m.
In both Fig.~\ref{fig:scene1_wb} and Fig.~\ref{fig:scene2_wb}, we see that there is a
displacement due to layover effect in the range direction ($x$-axis).
The first antenna reconstructs the target at $(-41,-31,0)$m.
The second antenna reconstructs the target at $(-48,-31,0)$m.

\begin{figure}[htpb]
\centering
{\includegraphics[width=3.5in]
          {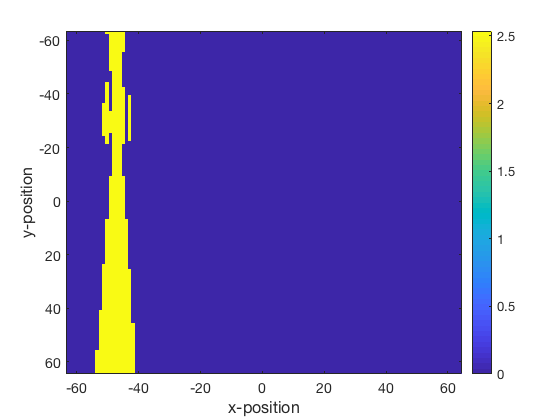}}
 \caption{The interferogram from wideband SAR reconstructed images.}
\label{fig:interferogram_wb}
\end{figure}

We next align the peaks in the two images and multiply the first image with the complex conjugate
of the second as in~\eqref{eq:SAR_WBint} to generate the interferogram.  The resulting
interferogram is shown in Fig.~\ref{fig:interferogram_wb}.

In order to reconstruct the height we use the set of
equations~\eqref{eq:InterIsoRangeDoppTopo4},~\eqref{eq:InterIsoRangeDoppTopo5},
and~\eqref{eq:InterIsoRangeDoppTopo6}.  The Doppler cone equation~\eqref{eq:InterIsoRangeDoppTopo5}
at zero-Doppler point $s_0^1$ gives us that the iso-Doppler contours are in the look-direction,
which in our scenario is parallel to the $x$-axis.  Thus, iso-Doppler contours have constant
$y$ value at the target's $y$ position.  Using this fact, we need only to compute the intersection
of iso-range contour~\eqref{eq:InterIsoRangeDoppTopo4} and interferometric phase
contour~\eqref{eq:InterIsoRangeDoppTopo6} fixing the $y$ position.  From Figs.~\ref{fig:scene1_wb}
and~\ref{fig:scene2_wb} we see that both targets are reconstructed at $y$ position of $-31$m.  Thus
we reconstruct the true target position using $y=-31$m.  For reconstruction, we sampled the height
in the interval $[1,100]$m at $0.5$m resolution.

\begin{figure}[htpb]
  \subfloat[][]{
     \includegraphics[width=0.52\textwidth]
     {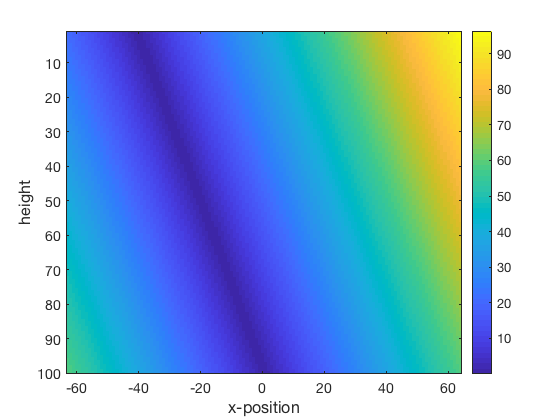}\label{fig:isoRange_wb}}
   \subfloat[][]{
     \includegraphics[width=0.52\textwidth]
     {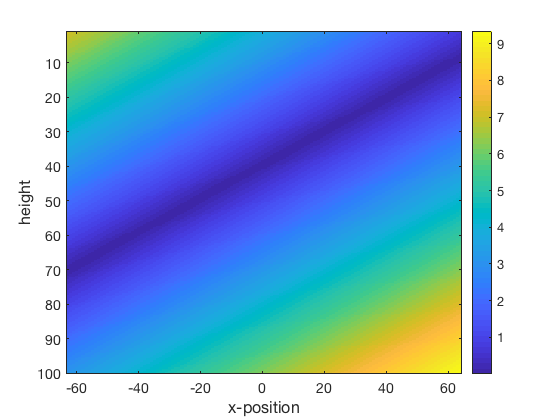}\label{fig:isoInterPhase_wb}
   }
   \caption{(a) Image of the magnitude of $|\z - \bgamma_1(s_0^1)| - R_1(\x, s_0^1)$ at $y=-31$m. The iso-range
   contour is indicated by dark blue area where the magnitude of $|\z - \bgamma_1(s_0^1)| - R_1(\x,
   s_0^1)$ is minimized.  (b) Image of the magnitude of $L_1(\brmn z, s_0^1)\cdot\brmn b  -
   \frac{c}{2\omega_0}\Phi^{WB}_{s_0}(\x)$ at $y=-31$m. The interferometric phase
   contour is indicated by dark blue area where the magnitude of $L_1(\brmn z, s_0^1)\cdot\brmn b
   - \frac{c}{2\omega_0}\Phi^{WB}_{s_0}(\x)$ is minimized.}
\end{figure}

Fig.~\ref{fig:isoRange_wb} shows the magnitude
image of $|\z - \bgamma_1(s_0^1)| - R_1(\x, s_0^1)$ at $y=-31$m.  Note that $R_1(\x,s_0^1)$ is the measured value derived from
the phase of the reconstructed image.  The dark blue area indicates the iso-range contour
where the magnitude of the difference is minimized.

Similarly, Fig.~\ref{fig:isoInterPhase_wb} shows the magnitude image of the difference
$L_1(\brmn z, s_0^1)\cdot\brmn b  - \frac{c}{2\omega_0}\Phi^{WB}_{s_0}(\x)$.  As before, the dark
blue area indicates the interferometric phase contour.

\begin{figure}[htpb]
\centering
{\includegraphics[width=3.5in]
          {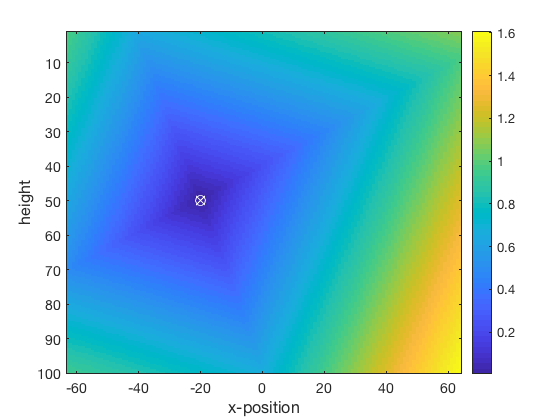}}
 \caption{Image of the intersection of the iso-range contour with the interfermetric phase
   contour at $y=-31$m.  The exact intersection is indicated by white `x'.  The true target position is indicated
 by white `o'.  The target is reconstructed at the correct position and height.}
\label{fig:intersect_wb}
\end{figure}

Combining the two images, Fig.~\ref{fig:intersect_wb} shows the intersection of the two contours
indicated by the dark blue area.  The white `x' in Fig.~\ref{fig:intersect_wb} indicates the exact
intersection computed and where the target is reconstructed.  The white `o' indicates the true
target position.  It is clear that the target is reconstructed at the correct position and height.
\subsection{Doppler-SAR}
\begin{figure}[htpb]
  \subfloat[][]{
    \includegraphics[width=0.52\textwidth]
    {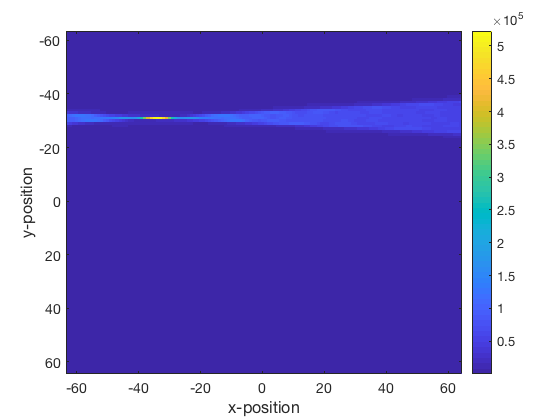}\label{fig:scene1_dop}
  }
  \subfloat[][]{
    \includegraphics[width=0.52\textwidth]
    {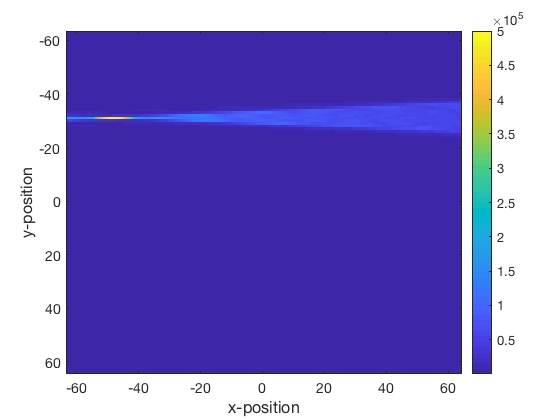}\label{fig:scene2_dop}}
  \caption{(a) Doppler-SAR reconstruction of the target located at $(-20,-31,50)$m using the first antenna assuming flat ground topography.  The target is
  reconstructed at $(-34,-31,0)$m.  (b) Doppler-SAR reconstruction of the target located at $(-20,-31,50)$m using the second antenna assuming flat ground topography.  The target is
  reconstructed at $(-48,-31,0)$m.}
\end{figure}

We proceed similar as in the wideband case for the Doppler-SAR case.  Figs.~\ref{fig:scene1_dop}
and~\ref{fig:scene2_dop} show the reconstructed image for Doppler-SAR for the first and second
antennas, respectively.
The first antenna reconstructs the target at $(-34,-31,0)$m and the second antenna at
$(-48,-31,0)$m.

\begin{figure}[htpb]
\centering
{\includegraphics[width=3.5in]
          {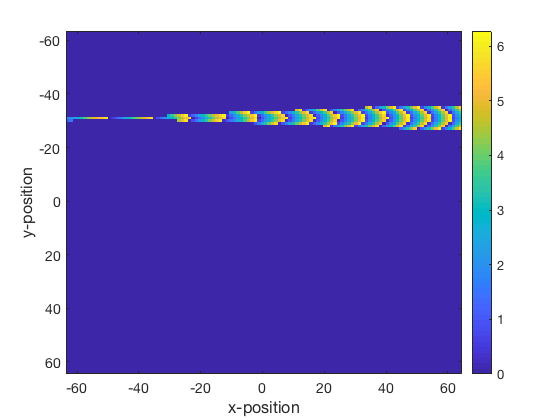}}
 \caption{The interferogram from Doppler-SAR reconstructed images.}
\label{fig:interferogram_dop}
\end{figure}

As in the wideband case, we align the peaks of the two images and multiply the
first image with the conjugate of the second image to form the interferogram of the Doppler
images. The resulting interferogram is shown in Fig.~\ref{fig:interferogram_dop}.

To reconstruct the height we use the set of equations given
in~\eqref{eq:InterIsoDoppTopo3},~\eqref{eq:InterIsoDoppTopo4}, and~\eqref{eq:InterIsoDoppTopo5}.
The zero-Doppler-rate points, $s_d^1$, is approximated by the end of the antenna's trajectories
farthest from the target position.  By~\eqref{eq:s_z_dop}, for a linear trajectory with constant
velocity, true zero-Doppler-rate point would be where
$\dot{\bgamma}_i(s_d^1)\perp\dot{\bgamma}_i^\perp(s_d^1)$.  Namely, where the look-direction is
parallel to the velocity vector.  The best estimate would be at a point in the trajectory farthest
away from the target location.

\begin{figure}[htpb]
  \centering
  \subfloat[][]{
     \includegraphics[width=0.52\textwidth]
     {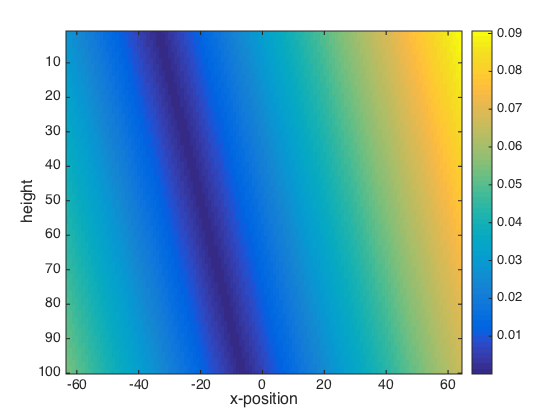}\label{fig:isoDopSurf}
   }
   \subfloat[][]{
     \includegraphics[width=0.52\textwidth]
     {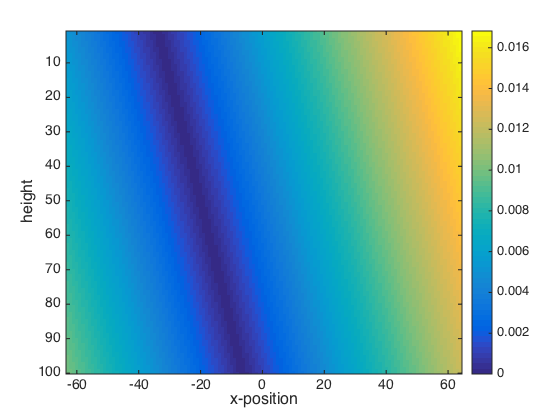}\label{fig:isoDopRateSurf}
   }

\subfloat[][]{
     \includegraphics[width=0.52\textwidth]
     {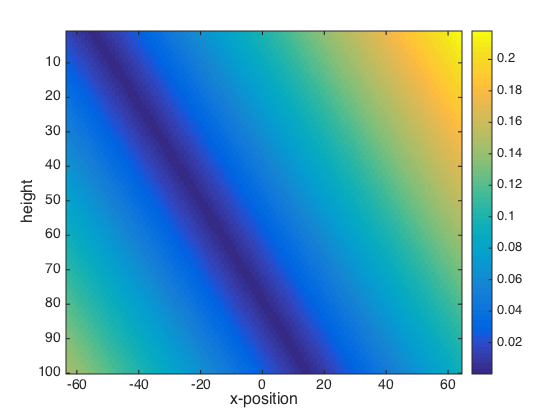}\label{fig:isoInterDopRateSurf}
   }
   \caption{(a) Image of the magnitude of $\widehat{(\z - \bgamma_1(s_d^1))}\cdot\dot{\bgamma}_1(s_d^1) - \frac{c}{\omega_0}f_1^d(\x, s_d^1)$ at $y=-31$m. The iso-Doppler
   contour is indicated by dark blue area where the magnitude of $\widehat{(\z -
     \bgamma_1(s_d^1))}\cdot\dot{\bgamma}_1(s_d^1) = \frac{c}{\omega_0}f_1^d(\x, s_d^1)$ is
   minimized.  (b) Image of the magnitude of $\widehat{(\z - \bgamma_1(s_d^1))}\cdot\ddot{\bgamma}_1(s_d^1)-\frac{\dot{\bgamma}_1(s_d^1)\cdot\dot{\bgamma}_1^\perp(s_d^1T_\phi)}{R_1(s_d^1,\z)} -\partial_s f_1^d(\x, s_d^1)$ at $y=-31$m. The iso-Doppler-rate
   contour is indicated by dark blue area where the magnitude of $\widehat{(\z -
     \bgamma_1(s_d^1))}\cdot\ddot{\bgamma}_1(s_d^1)-\frac{\dot{\bgamma}_1(s_d^1)\cdot\dot{\bgamma}_1^\perp(s_d^1T_\phi)}{R_1(s_d^1,\z)}
   - \partial_s f_1^d(\x, s_d^1)$ is minimized.  (c) Image of the magnitude of $\L_1(\z,s_d^1)\cdot
   \v + \frac{\b_1^\perp \cdot \dot{\bgamma}_2(s_d^2)}{R_1(\z,s_d^1)}
   +\frac{c}{2s_d^1T_\phi\omega_0}\Phi^{UNB}_{s_d}(\x)$ at $y=-31$m. The interferometric Doppler-rate
   contour is indicated by dark blue area where the magnitude of $\L_1(\z,s_d^1)\cdot \v +
   \frac{\b_1^\perp \cdot \dot{\bgamma}_2(s_d^2)}{R_1(\z,s_d^1)}
   +\frac{c}{2s_d^1T_\phi\omega_0}\Phi^{UNB}_{s_d}(\x)$ is minimized.}

\end{figure}

Fig.~\ref{fig:isoDopSurf} illustrates the iso-Doppler surface at $y=-31$m, which is the
$y$-position where the target position is reconstructed and the true target's $y$ position.  Notice
that both images reconstruct the scatterer at the correct $y$ position.  The iso-Doppler contour is given by the
dark blue area as before.

Similarly, Figs.~\ref{fig:isoDopRateSurf} and~\ref{fig:isoInterDopRateSurf} illustrate the
iso-Doppler-rate and interferometric Doppler-rate surfaces, respectively at $y=-31$m.



\begin{figure}[htpb]
\centering
{\includegraphics[width=3.5in]
          {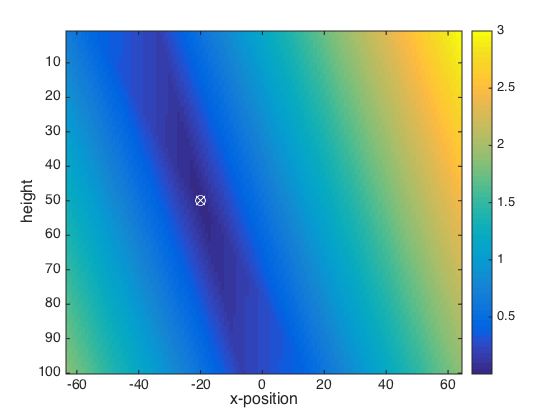}}
 \caption{Image of the intersection of the iso-Doppler, iso-Doppler-rate and interferometric
   Doppler-rate contours at $y=-31$m.  The intersection is indicated by white `x'.  The true target position is indicated
 by white `o'.  The target is reconstructed at the correct position and height.}
\label{fig:intersect_dop}
\end{figure}

Fig.~\ref{fig:intersect_dop} combines Figs.~\ref{fig:isoDopSurf},~\ref{fig:isoDopRateSurf},
and~\ref{fig:isoInterDopRateSurf}.  The intersection of the three contours is indicated by white
`x'.  The white `o' shows the true target location.
Clearly, the target is reconstructed at the correct position and height.

\section{Conclusions}\label{sec:conclusion}
We present a novel radar interferometry based on Doppler-SAR imaging paradigm.  Doppler-SAR uses
single frequency transmitted waveforms.
It has several advantages over
conventional SAR including simpler, inexpensive hardware, high SNR and long effective range of
operation, and is suitable for use in passive radar applications.

We derived the interferometric phase relationship for Doppler-SAR.  Doppler-SAR interferometric phase
depends on the difference in the velocity of the antennas as opposed to the range difference
observed in
wideband SAR.  Thus, in Doppler-SAR interferometry, one can
reconstruct the ground topography even with the same look-direction from both
antennas so long as their velocities are different).
Furthermore, we showed that the true target position is determined by the intersection
of iso-Doppler,
iso-Doppler-rate, and interferometric Doppler-rate surfaces.  This is different from
conventional wideband SAR in that the surfaces that determine the true target position are iso-range,
iso-Doppler, and interferometric Doppler-rate surfaces.

We presented numerical simulations for a single point scatterer
using
two antennas moving in linear trajectories
to verify our interferometric
method.  We also conduct conventional wideband SAR interferometric reconstruction as a comparison.
We show that both wideband SAR and Doppler-SAR
interferometry is able to accurately reconstruct the target location.
Thus, our numerical simulations show that Doppler-SAR interferometry retains the accuracy of
conventional SAR interferometry while having the advantage that
Doppler-SAR affords.

In the future, we will analyze the sensitivity of height estimation with respect to other
observables and parameters.

\section*{Acknowledgement}
This material is based upon work supported by the Air Force Office of Scientific Research (AFOSR) under award number FA9550-16-1-0234,
and by the National Science Foundation (NSF) under Grant No. CCF-1421496.
\newpage

\appendix
\section{Approximations}\label{app:ff_DL}
\subsection{Far-field approximation}
Let $\brmn x$ and $\brmn y$ be two vectors such that $|\brmn x|\gg|\brmn y|$. Then, by using Taylor series expansion we can make the following approximation:
\begin{eqnarray}
|\brmn x-\brmn y|&=&\sqrt{(x_1-y_1)^2+(x_2-y_2)^2+(x_3-y_3)^2},\\
&=&\sqrt{|\brmn x|^2-2(x_1y_1+x_2y_2+x_3y_3)+|\brmn y|^2},\\
&=&|\brmn x|\sqrt{1-\frac{2(\brmn x\cdot\brmn y)}{|\brmn x|^2}+\frac{|\brmn y|^2}{|\brmn x|^2}},\\
&\approx&|\brmn x|\left(1-\frac{1}{2}\frac{2\brmn x\cdot\brmn y}{|\brmn x|}\right),\\
&\approx&|\brmn x|-\hat{\brmn x}\cdot\brmn y
\end{eqnarray}
where $\hat{\brmn x}$ is the unit vector $\hat{\brmn x}=\frac{\brmn x}{|\brmn x|}$.
\subsection{Approximation of look-direction under far-field assumption}
Let $\widehat{(\brmn x-\bgamma(s))}$ denote a look direction where $\brmn x=\brmn y+\brmn z$ and $|\brmn y-\bgamma(s)|\gg |\brmn z|$. Then by using far field expansion we can write
\begin{eqnarray}
\fl\qquad\widehat{(\brmn x-\bgamma(s))}&=&\frac{\brmn x-\bgamma(s)}{|\brmn x-\bgamma(s)|}\\
&=&\frac{\brmn y+\brmn z-\bgamma(s)}{|\brmn y+\brmn z-\bgamma(s)|},\\
&\approx&\frac{\brmn y-\bgamma(s)}{|\brmn y-\bgamma(s)|+\widehat{(\brmn y-\bgamma(s))}\cdot\brmn z}+\frac{\brmn z}{|\brmn y-\bgamma(s)|
+\widehat{(\brmn y-\bgamma(s))}\cdot\brmn z},\\
&\approx&(|\brmn y-\bgamma(s)|-\widehat{(\brmn y-\bgamma(s))}\cdot\brmn z)\frac{\brmn
          y-\bgamma(s)}{|\brmn y-\bgamma(s)|^2}\cr
          & &\qquad\qquad\ +\frac{(|\brmn y-\bgamma(s)|-\widehat{(\brmn y-\bgamma(s))}\cdot\brmn z)\brmn z}{|\brmn y-\bgamma(s)|^2},\\
&\approx&\widehat{(\brmn y-\bgamma(s))}+\frac{\brmn z-\widehat{(\brmn y-\bgamma(s))}\left[\widehat{(\brmn y-\bgamma(s))}\cdot\brmn z\right]}{|\brmn y-\bgamma(s)|},\\
&\approx&\widehat{(\brmn y-\bgamma(s))}+\frac{\brmn z^\perp}{|\brmn y-\bgamma(s)|}
\end{eqnarray}
where $\brmn z^\perp$ is the transverse $\brmn z$, i.e. projection of $\brmn z$ onto the plane whose normal vector is along the look direction $\widehat{(\bgamma(s)-\brmn y)}$. Therefore, difference of look directions is given by:
\begin{equation}
\widehat{(\brmn x-\bgamma(s))}-\widehat{(\brmn y-\bgamma(s))}\approx\frac{\brmn z^\perp}{|\bgamma(s)-\brmn y|}
\end{equation}
where $\brmn x=\brmn y+\brmn z$.
\section*{References}
\bibliography{dopperinsar_v2}
\end{document}